# A dynamic reconstruction and motion estimation framework for cardiorespiratory motion-resolved real-time volumetric MR imaging (DREME-MR)

Hua-Chieh Shao[1], Xiaoxue Qian[1], Guoping Xu[1], Can Wu[2], Ricardo Otazo[2], Jie Deng[1], and You Zhang[1]

*[1] The Medical Artificial Intelligence and Automation (MAIA) Laboratory*
*[1] Department of Radiation Oncology, University of Texas Southwestern Medical Center, Dallas, TX 75390, USA*
*[2] Department of Medical Physics, Memorial Sloan Kettering Cancer Center, New York, NY 10065, USA*

Corresponding address:

You Zhang
Department of Radiation Oncology
University of Texas Southwestern Medical Center
2280 Inwood Road
Dallas, TX 75390
Email: You.Zhang@UTSouthwestern.edu
Tel: (214) 645-2699

## Abstract

**Objective:** Based on a 3D pre-treatment magnetic resonance (MR) scan, we developed DREME-MR to jointly reconstruct the reference patient anatomy and solve a data-driven, patient-specific cardiorespiratory motion model. Via a motion encoder simultaneously learned during the reconstruction, DREME-MR further enables real-time volumetric MR imaging and cardiorespiratory motion tracking with minimal intra-treatment k-space data.

**Approach:** DREME-MR integrates dynamic MRI reconstruction and real-time MR imaging into a unified, dual-task learning framework. From a 3D radial-spoke-based pre-treatment MR scan, DREME-MR uses spatiotemporal implicit-neural-representation (INR) to reconstruct pre-treatment dynamic volumetric MR images (learning task 1). The INR-based reconstruction takes a joint image reconstruction and deformable registration approach, yielding a reference anatomy and a corresponding cardiorespiratory motion model. The motion model adopts a low-rank, multi-resolution representation to approximate motion fields as products of motion coefficients and motion basis components (MBCs). Via a progressive, frequency-guided strategy, DREME-MR decouples cardiac MBCs from respiratory MBCs to resolve the two distinct motion modes. Simultaneously with the pre-treatment dynamic MRI reconstruction, DREME-MR also trains a multilayer perceptron (MLP)-based motion encoder to infer cardiorespiratory motion coefficients directly from the raw k-space data (learning task 2), allowing real-time, intra-treatment volumetric MR imaging and motion tracking with minimal k-space data (20-30 spokes) acquired after the pre-treatment MRI scan.

**Main results:** Evaluated using data from a digital phantom (XCAT) and a human scan, DREME-MR solves real-time 3D cardiorespiratory motion with a latency of < 165 ms (= 150-ms data acquisition + 15-ms inference time), fulfilling the temporal constraint of real-time imaging. The XCAT study achieves mean(±S.D.) center-of-mass tracking errors of 0.73±0.38mm for a lung tumor and 1.69±1.12mm for the left ventricle. The human study shows good motion correlations (liver: 0.96; left ventricle: 0.65) between DREME-MR-solved motion and extracted surrogate signals.

**Significance:** DREME-MR allows real-time 3D MRI and cardiorespiratory motion tracking with low latency, advancing intra-treatment MR-guided adaptive radiotherapy, including real-time multileaf collimator (MLC) tracking.

## 1. Introduction

Magnetic resonance imaging (MRI) offers high soft-tissue contrast for improved anatomical visualization and morphological delineation, without exposing patients to ionizing radiation. In addition to structural information, MRI can provide functional and molecular information to help with disease diagnosis and prognosis (Bartsch et al. 2006). Due to advances in MRI technology, computational algorithms, and cost reduction (Harisinghani et al. 2019), MRI has been introduced into and gradually integrated into the clinical workflow of radiotherapy (Martin et al. 2021, Otazo et al. 2021), providing image guidance in MRI-only treatment planning (Owrangi et al. 2018, Greer et al. 2019) and MRI-guided radiotherapy delivery (Corradini et al. 2019, Hall et al. 2019, Keall et al. 2022). Radiotherapy treats patients with tumors using high-energy X-rays or particles. An effective treatment demands precise and accurate three-dimensional (3D) dose distributions conformal to treatment targets to achieve desirable tumor control while avoiding radiotoxicity to surrounding healthy tissues (Chandra et al. 2021). However, for thoracic and abdominal patients, variations in target position and shape caused by anatomical motion remain a major source of uncertainties in precise dose delivery, thereby potentially compromising treatment efficacy and outcomes (Seppenwoolde et al. 2002, Bertholet et al. 2016).

Standard patient motion management practices typically rely on 4D-MRI to analyze and characterize motion patterns (Stemkens et al. 2018), upon which personalized motion management approaches can be chosen (e.g., breath-hold or respiratory gating) to minimize tumor localization uncertainties (Paganelli et al. 2018, Ball et al. 2022). 4D-MRI reconstruction usually uses external/internal motion surrogate signals to sort acquired k-space data into pre-defined respiratory motion states (i.e., bins), and then volumetric MR images of different motion bins are individually reconstructed (Stemkens et al. 2018, Menchon-Lara et al. 2019, Rajiah et al. 2023). To compensate for incomplete measurements of each motion state, 4D-MRI repeatedly measures patient movement, thus prolonging scan time and increasing the likelihood of motion variation-related artifacts. Motion sorting assumes anatomical motion is perfectly reproducible, which is often inaccurate, as patients frequently exhibit irregular motion (e.g., breathing frequency/amplitude variations and baseline drifts). As a result, motion sorting may significantly degrade image quality. More importantly, 4D-MRI is unable to capture irregular motion, which can provide important information for motion management decisions and patient functional assessments. *Dynamic volumetric MRI*, on the other hand, offers a solution to address the above limitations. Dynamic volumetric MRI here refers to retrospective reconstruction of 3D MR images with much higher temporal resolution to capture transient events (Nayak 2019), without using external/internal surrogate signals for motion sorting. Each volume of dynamic MRI is reconstructed based on very limited k-space data (tens of k-space spokes, for instance) to eliminate the motion within. Accordingly, it eliminates the need for motion sorting and thus related artifacts. However, the reconstruction of dynamic volumetric MRI is a highly ill-posed spatiotemporal inverse problem, as the volumetric information becomes severely undersampled for each MR image. To address the extreme undersampling issue, traditional dynamic MRI reconstruction methods exploit spatiotemporal redundancy and correlations within MR acquisitions, combined with compressed sensing and parallel imaging, to reconstruct a dynamic sequence of MRIs (Tsao et al. 2003, Feng et al. 2016, Ong et al. 2020,

Ravishankar et al. 2020, Murray et al. 2024). More recently, data-driven and learning-based approaches have been proposed to remove image noise and aliasing artifacts in undersampled MR images (Ravishankar and Bresler 2011, Liang et al. 2020, Singh et al. 2023). Traditionally, dynamic MRI is limited to 2D imaging, due to the complexity involved in spatiotemporal reconstruction and optimization. Since organ and tumor motion in the thorax and abdomen exhibits complex 3D dynamics (Langen and Jones 2001), volumetric imaging is highly desirable for accurate 3D motion estimation and characterization. Learning-based approaches can handle 3D reconstruction more effectively but require large datasets to train the models, which are often not available. Recent studies such as MR-MOTUS (Huttinga et al. 2020, Huttinga et al. 2021), Extreme MRI (Ong et al. 2020), and STINR-MR (Shao et al. 2024) focus on 3D imaging and reconstruction based on a single MR scan. They use the full dataset to leverage the spatiotemporal correlation between sequential dynamic volumes for collective reconstruction, addressing the undersampling problem.

While dynamic volumetric MRI provides rich and valuable motion information for personalized motion management, the collective reconstruction approach of algorithms like MR-MOTUS, Extreme MRI, and STINR-MR limits its ability to fully address the challenge of motion-related uncertainties during radiation treatment, as it needs to use the entire acquisition dataset for time-consuming, spatiotemporally-correlated or motion-compensated reconstruction. The instantaneous motion variations that occur during the treatment require *real-time MR imaging* (Bertholet et al. 2019, Nayak et al. 2022, Lombardo et al. 2024) to capture anatomical information with sub-second latency during radiation delivery, thereby enabling real-time treatment verification and adaptation (Keall et al. 2019, McNair and Buijs 2019, Keall et al. 2025). To achieve such real-time imaging and motion tracking, stringent constraints are imposed on system latency. Due to fast anatomical motion, it has been suggested that the temporal latency should be limited to 500 milliseconds (ms) for respiratory motion (Keall et al. 2021) and 200 ms for cardiac motion (Campbell-Washburn et al. 2017), which includes both image acquisition and reconstruction time. Because the sampling rate of MRI acquisition is inherently slow, only limited anatomical information is sampled for volumetric reconstruction within such a short time interval. The reconstruction also needs to be extremely fast, preventing the joint use of previously acquired data for time-consuming spatiotemporal reconstructions, as done for dynamic volumetric MRI. Thus, achieving real-time imaging requires fast MR acquisition, efficient reconstruction and tracking algorithms, and significant computational power.

With the recent development of deep learning (DL) and high-speed GPU computing, many DL-based approaches have been proposed for real-time imaging and motion tracking in MRI-guided radiotherapy. DL methods for MRI-based real-time imaging or motion tracking can be broadly categorized into reconstruction-based and registration-based approaches. Reconstruction approaches either directly generate high-quality MR images from undersampled k-space acquisitions (Zhu et al. 2018) or formulate the reconstruction problem as a de-aliasing/de-noising process in the image domain (Liu et al. 2022). For example, Schlemper et al. (Schlemper et al. 2018) proposed a cascaded DL model to reconstruct 2D cardiac MR images from aliased input images, alternating between convolutional neural networks and data consistency layers to resemble iterative de-aliasing algorithms. Yang et al. (Yang et al. 2018) developed a conditional generative adversarial network for compressed sensing MRI reconstruction. They incorporated perceptual loss alongside adversarial learning to enhance image details, with an inference time of 5 ms for a 2D brain MR image. Huang et al. (Huang et al. 2022) introduced a Swin transformer-based DL model for fast 2D MRI reconstruction, utilizing shifted windows multi-head self-attention mechanism to de-alias zero-filled images. However, these methods, although allowing fast reconstruction, are mostly limited to 2D. Moreover, additional segmentation steps are necessary for these reconstruction-based approaches to locate moving targets, which can introduce further localization uncertainties and increase the system latency.

To achieve 3D imaging and target localization under severely undersampled scenarios, registration-based DL approaches were proposed (Terpstra et al. 2020, Terpstra et al. 2021, Shao et al. 2022, Hunt et al. 2023, Wei et al. 2023, Lombardo et al. 2024). In particular, Terpstra et al. (Terpstra et al. 2021) proposed a DL model (TEMPEST) that estimates a 3D motion field between a pair of high-quality static MR volume and undersampled dynamic (moving) MR volume under 200-ms latency (including the time of MR acquisition), using a multi-resolution pyramid registration scheme. They achieved high-quality motion fields with a < 2-mm registration accuracy for the cases of a 366-fold undersampling ratio. However, TEMPEST was based on supervised learning, requiring 'ground-truth' 3D motion fields as training labels. Since the 3D motion field labels were solved by other approaches, and physiologically realistic, 'ground-truth' motion fields are hard to obtain, the registration errors in the label motion fields may propagate to the DL model, leading to intrinsic biases of the model. To address these potential biases, Shao et al. (Shao et al. 2022) developed an unsupervised DL model (KS-RegNet) for real-time motion estimation, based on the Voxelmorph architecture (Balakrishnan et al. 2019). The model training was driven by a k-space data consistency loss matching re-projected k-space data of registered images with undersampled k-space acquisitions, thus avoiding the need for 'ground-truth' motion fields. They achieved a localization accuracy of < 2 mm for an 80-fold undersampling ratio, under ~600-ms latency (including the time of MR acquisition). Due to the use of non-uniform Fourier transformation in KS-RegNet, the overall latency is relatively long. Wei et al. (Wei et al. 2023) proposed a similar unsupervised approach that registers a prior 3D MRI to onboard coronal 2D MRIs to generate new 3D real-time MRIs. They achieved a localization error < 2.6 mm under 100-ms latency (excluding the time for MR acquisition). However, these registration-based approaches typically incorporate patient-specific prior information (e.g., patient anatomy from a different scan, previously-derived motion models, and/or motion surrogate signals) into the motion estimation. While this can enhance localization accuracy, it may introduce biases in motion estimation, as the patient anatomy, imaging contrast, and motion can vary during the course of treatment. Moreover, these DL-based methods may suffer from generalizability and robustness issues when applied to out-of-distribution data, as DL model training usually requires large MR datasets, which are limited in availability.

In addition to the above DL-based methods, Huttinga et al. (Huttinga et al. 2022) extended their dynamic MRI reconstruction framework, MR-MOTUS (Huttinga et al. 2020, Huttinga et al. 2021), for real-time imaging to solve non-rigid 3D respiratory motion fields. The framework divides real-time motion estimation into an offline preparation phase and a real-time online phase. During the preparation phase, the modified real-time MR-MOTUS method uses an iterative reconstruction algorithm with a B-spline-based motion model to solve a 10-phase 4D-MRI, based on a 10-minute MR acquisition. When the method is deployed to real-time MRI, it leverages the solved anatomy and motion model from the preparation phase to solve a real-time motion field in a single iteration, using a 67-ms MR acquisition. They achieved a total latency of 170 ms. However, the reference anatomy was independently reconstructed from their motion model without motion-compensated reconstruction, thus potentially causing inconsistency and incoherence between them (Shao et al. 2024). Recently, Wu et al. introduced MRSIGMA (Wu et al. 2023), a similar framework that uses XD-GRASP (Feng et al. 2016) in an offline dictionary-learning phase to create a 10-phase 4D motion dictionary that uniquely associates MR motion signatures with 4D motion states. During real-time imaging, MRSIGMA performs signature matching to determine the corresponding motion states. Since both approaches require motion-sorted 4D-MRI reconstruction in the preparation phase, they suffer from the aforementioned 4D-MRI limitations. Similar to most 4D-MRI-based works, their framework focuses on respiratory motion only, without resolving the cardiac motion. However, studies have shown a correlation between cardiac dose and radiotherapy-associated cardiac toxicity in lung and breast cancer patients (Vivekanandan et al. 2017, Atkins et al. 2019, Omidi et al. 2023), highlighting the need for cardiorespiratory motion models for cardiac dose mapping and MR imaging to improve patient safety.

Furthermore, the growing use of radiotherapy in cardiac radioablation to treat ventricular tachycardia also underscores the importance of accurate cardiorespiratory motion models for heart patients (van der Ree et al. 2020, Lydiard et al. 2021).

To address the above challenges, in this work, we propose a dual-task learning framework, called dynamic reconstruction and motion estimation for MR (DREME-MR), that integrates dynamic volumetric MRI reconstruction into a real-time imaging framework. DREME-MR combines two learning objectives into one training session: (1) to reconstruct a sequence of dynamic volumetric MRIs from a pre-treatment 3D MR scan to acquire an up-to-date patient anatomy and patient-specific motion model, via a motion-compensated framework that simultaneously optimizes the image and the motion model; and (2) during dynamic MRI reconstruction, to concurrently train a neural network-based motion encoder capable of estimating motion states for subsequent real-time imaging and motion tracking, based on minimal new k-space data acquired in real time. By the first learning objective, DREME-MR addresses the ill-posed spatiotemporal inverse problem of dynamic volumetric MR reconstruction by utilizing a joint reconstruction and deformable registration approach on all motion-contained data from a pre-treatment MR scan. Following a 'one-shot' learning strategy, DREME-MR does not require large population-based datasets for pre-training, as it uses only the k-space data acquired from a patient-specific pre-treatment MR scan. The temporal proximity between model training and its deployment for real-time imaging helps minimize the risk of data distribution shift and enhances the applicability of the model to patient-specific anatomy/motion. In addition, given DREME-MR's high spatiotemporal resolution (3 mm and 100-150 ms) and the growing need to resolve cardiac motion in radiotherapy (van der Ree et al. 2020, Lydiard et al. 2021), we extended DREME-MR's motion model to include cardiac motion. Based on the frequency and motion region differences between cardiac and respiratory motion, we developed a frequency-guided training scheme and decoupled coordinate systems to facilitate DREME-MR to solve and optimize distinct cardiac and respiratory motion modes. DREME-MR was validated using a digital phantom-based simulation study and a human subject study. Additionally, it was compared with principal component analysis-based motion modeling and reconstruction algorithms and two other dynamic MRI reconstruction methods.

## 2. Theory

### 2.1 Problem formulation

Consider a pre-treatment 3D MR scan covering the thoracic-abdominal region of a subject. The aim of dynamic MRI reconstruction in this work is to solve a sequence of dynamic images that visualize cardiorespiratory-induced anatomical motion for analysis and treatment guidance. Specifically, k-space of the moving anatomy is continuously sampled using a golden-mean radial trajectory (Chan et al. 2009, Feng 2022), where each readout line is an oriented radial spoke that diagonally traverses the 3D k-space and passes through the origin. We want to note here that the DREME-MR algorithm is not limited to the 3D radial trajectory and can be readily applied to other trajectories like the stack-of-stars (Chandarana et al. 2011, Feng 2022). For 3D radial spokes, the readout orientations are calculated according to the multidimensional golden-mean algorithm (Winkelmann et al. 2007, Chan et al. 2009). Because of these properties, the golden-mean radial trajectory has been demonstrated to be robust to motion-related artifacts (Chan et al. 2009, Hamilton et al. 2017, Feng 2022). From these 3D radial spokes, a motion-resolved sequence of MRI frames can be reconstructed. A frame here is defined as an MR volume with sufficient temporal resolution such that the anatomical state captured by each frame can be considered static with negligible movement. In this study, each 3D MR scan lasts approximately 4 minutes, from which a dynamic sequence consisting of 1,000-2,000 frames of volumetric MRIs can be reconstructed, equivalent to a temporal resolution of ~100-200 ms, which is sufficient to resolve cardiorespiratory motion (Campbell-Washburn et al. 2017, Keall et al. 2021). Based on a pulse sequence with a repetition time (TR) of $\lesssim 5$ ms, each frame contains around 20-40 spokes, corresponding to an undersampling ratio of ~1,400-2,700 (estimated by assuming uniform angular sampling in the radial and azimuthal angles and an MRI volume of 150×150×150 voxels). Additionally, the k-space data are acquired by multi-channel phased array coils (~20 coils), thus providing localized spatial information to accelerate MR scan and facilitate MRI reconstruction via sensitivity spatial encoding. Overall, the MR scan comprises an order of $\mathcal{O}(10^6)$ k-space sampling points for each coil.

### 2.2 Dynamic MRI reconstruction algorithm and cardiorespiratory motion model

As aforementioned in the introduction, dynamic MRI reconstruction is a highly ill-posed spatiotemporal inverse problem typically involving more than $\mathcal{O}(10^9)$ unknowns (Huttinga et al. 2021). To condition the reconstruction process, we adopted a joint reconstruction and registration approach with a low-rank motion model, based on the following two observations: (a) Anatomies captured at different frames are highly correlated. Accordingly, we exploited this temporal correlation, assuming that voxel intensity variations across frames can be accounted for by anatomical motion. We therefore hypothesized that there exists a reference anatomy $\boldsymbol{I}_{ref}(\boldsymbol{x})$, and each frame $\boldsymbol{I}(\boldsymbol{x},t)$ in the dynamic sequence can be obtained from a deformable registration of $\boldsymbol{I}_{ref}(\boldsymbol{x})$:

$$\boldsymbol{I}(\boldsymbol{x},t) = \boldsymbol{I}_{ref}(\boldsymbol{x} + \boldsymbol{d}(\boldsymbol{x},t)), \tag{1}$$

where $\boldsymbol{x}$ and $t$ respectively denote the voxel coordinates of the reconstruction volume and the frame index of the dynamic sequence, and $\boldsymbol{d}(\boldsymbol{x},t)$ denotes the dynamic deformation vector field (DVF) that represents the anatomical motion at frame $t$. This registration-based approach assumes the MR signals acquired at different time points satisfy the steady-state condition and local-spin conservation (Huttinga et al. 2020), thus excluding cases involving image contrast variations such as dynamic contrast-enhanced MRI (Padhani 2002) or non-stationary state MR acquisition (Tippareddy et al. 2021). Note that Eq. (1) does not include the Jacobian determinant of the DVF to reduce computational complexity, thereby not accounting for voxel volume changes. This simplification is typically used in registration-based methods, as incorporating the Jacobian determinant is expected to provide little improvement in accuracy, as most biological tissues are nearly incompressible during anatomical motion. (b) The second observation is that anatomical motion, especially heartbeat and respiration, exhibits spatially correlated and temporally quasi-cyclic motion patterns. This spatiotemporal motion correlation indicates that the time-varying motion field $\boldsymbol{d}(\boldsymbol{x},t)$ can be well approximated in a low-dimensional function space (Zhang et al. 2007, Li et al. 2011, Stemkens et al. 2016). Therefore, to further alleviate the ill-posedness of the inverse problem, a low-rank motion model was employed, in which the time-dependent motion field $\boldsymbol{d}(\boldsymbol{x},t)$ is decomposed into products of spatial and temporal components:

$$\boldsymbol{d}(\boldsymbol{x},t) = \sum_{i=1}^{L} \boldsymbol{w}_i(t) \times \boldsymbol{e}_i(\boldsymbol{x}), \tag{2}$$

where $L$ represents the number of levels in the decomposition. The spatial components $\{\boldsymbol{e}_i(\boldsymbol{x})\}_{i=1}^{L}$ can be considered as a basis set that spans a Hilbert subspace, and all motion states within the dynamic sequence can be accounted for by scaling the spatial components via the corresponding temporal components $\{\boldsymbol{w}_i(t)\}_{i=1}^{L}$ representing the motion coefficients at $t$. Because of this property, $\{\boldsymbol{e}_i(\boldsymbol{x})\}_{i=1}^{L}$ and $\{\boldsymbol{w}_i(t)\}_{i=1}^{L}$ are called motion basis components (MBCs) and MBC scores, respectively, in this work. With the above regularization approaches, the number of unknowns reduces to ~$\mathcal{O}(10^7)$.

The number of levels $L$ in the low-rank motion model Eq. (2) depends on the complexity of the anatomic motion of interest. Three levels ($L$ = 3) were proved sufficient for modeling respiratory motion (Li et al. 2011). Since cardiac and respiratory motions span wide and disparate spatial and temporal scales, we used separate low-rank MBCs for cardiac and respiratory motions and exploited inherent differences between the two motion characteristics. Since cardiac motion is more spatially localized than respiratory motion, we decoupled the two spatial scales by introducing an independent, localized cardiac coordinate system enclosing the heart. In order words, in terms of the motion model, the motion space is spanned by global respiratory $\boldsymbol{e}_i^r(\boldsymbol{x})$ and local cardiac $\boldsymbol{e}_i^c(\boldsymbol{x})$ MBCs. Furthermore, the deformable registration in Eq. (1) is performed in a sequential order that the cardiac deformable registration is performed first, followed by the respiratory deformable motion:

$$\boldsymbol{I}'(\boldsymbol{x},t) = \boldsymbol{I}_{ref}\big(\boldsymbol{x} + \boldsymbol{d}_c(\boldsymbol{x},t)\big) \quad \text{then} \quad \boldsymbol{I}(\boldsymbol{x},t) = \boldsymbol{I}'\big(\boldsymbol{x} + \boldsymbol{d}_r(\boldsymbol{x},t)\big)\ , \tag{3}$$

where $\boldsymbol{d}_r$ and $\boldsymbol{d}_c$ respectively denote the respiratory and cardiac DVFs, and $\boldsymbol{I}'(\boldsymbol{x},t)$ is an intermediate anatomy with only anatomical motion around the heart. In addition to the spatial decoupling, a frequency-guided regularization (see Sec. 2.4 for details) is utilized to further separate the motions in frequency domain. Note that the sequential deformable registration in Eq. (3) does not imply a temporal order between cardiac and respiratory motions, nor does it assume $\boldsymbol{d}_r$ and $\boldsymbol{d}_c$ are statistically independent. Importantly, our approach explicitly allows for dependency between the two motion components to account for their inherent entanglement, as the second (respiratory) motion field typically depends on the first (cardiac) motion field to accurately model the composite cardiorespiratory motion. Additionally, no orthogonality condition is imposed on the respiratory MBCs $\boldsymbol{e}_i^r(\boldsymbol{x})$, as we found it impedes model learning efficiency.

With the above strategies, the dynamic reconstruction problem is solved by the following optimization problem combining a k-space data consistency term with a regularization term $R$:

$$\hat{\boldsymbol{I}}(\boldsymbol{x},t) = \underset{\boldsymbol{I}_{ref}(\boldsymbol{x}),\boldsymbol{w}_i(t),\boldsymbol{e}_i(\boldsymbol{x})}{\operatorname{argmin}} \mathbb{E}[\|F[\boldsymbol{I}(\boldsymbol{x},t)] - \boldsymbol{s}(\boldsymbol{k},t)\|_1] + \lambda R[\boldsymbol{I}(\boldsymbol{x},t)], \tag{4}$$

where $\mathbb{E}$ denotes the expectation value averaged over sampled k-space points $\boldsymbol{k}$ in the scan, $\|\cdot\|_1$ is the L1 norm, $F$ is the operator combining the coil sensitivity encoding and non-uniform fast Fourier transform (NUFFT), $\boldsymbol{s}(\boldsymbol{k},t)$ denotes the acquired k-space signals at frame $t$, and $\lambda$ represents the weighting factor for the regularization term $R$. The first term on the right-hand side of Eq. (4) is the k-space data consistency loss that enforces the reconstructed k-space data to match the acquired k-space signals $\boldsymbol{s}(\boldsymbol{k},t)$. The second term is an image smoothness and motion model regularization term in the optimization process to mitigate the undersampled reconstruction challenge (see Sec. 2.4 for details). The L1 norm was used to quantify k-space data consistency, as it empirically led to faster convergence during model training compared to the L2 norm.

After completing the dynamic reconstruction, DREME-MR yields an up-to-date anatomy $\boldsymbol{I}_{ref}(\boldsymbol{x})$ and the corresponding cardiorespiratory motion model (i.e., $\boldsymbol{w}_i(t)$ and $\boldsymbol{e}_i(\boldsymbol{x})$). As DREME-MR was designed with the capability to infer the temporal motion amplitudes $\boldsymbol{w}_i(t)$ from limited-sampling k-space acquisition $\boldsymbol{s}(\boldsymbol{k},t)$ (see Sec. 2.2 for details), it can be directly applied during subsequent treatment delivery after the pre-treatment scan to achieve real-time volumetric imaging and motion tracking.

## 3. Materials and Methods

### 3.1 DREME-MR framework and dual-task learning

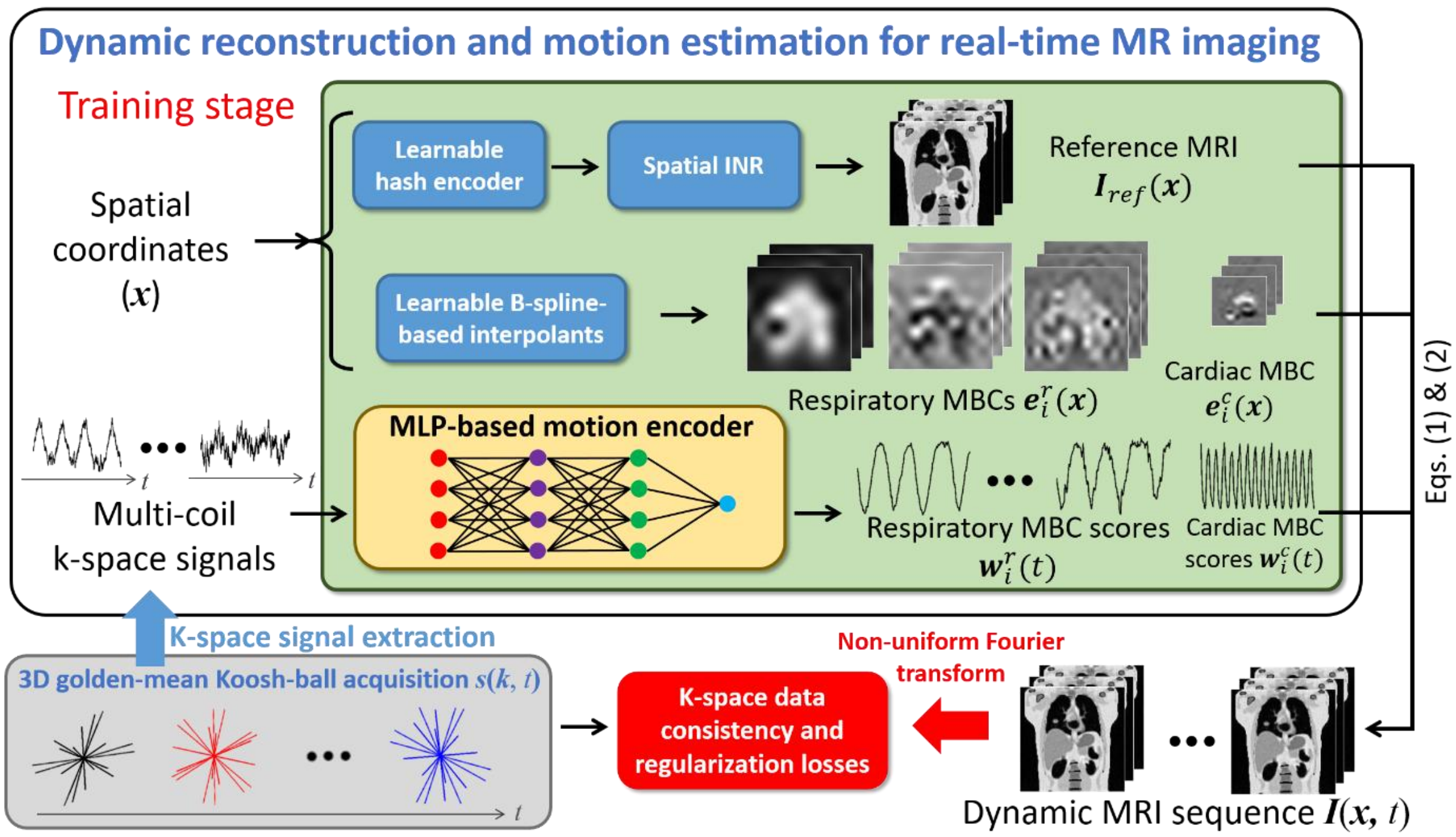

**Figure 1.** Overview of the DREME-MR framework and the dual-task learning strategy. DREME-MR simultaneously reconstructs a dynamic sequence of volumetric MR images (learning task 1) and trains a multilayer perceptron (MLP)-based motion encoder for real-time imaging and motion tracking (learning task 2), using a pre-treatment MR scan. DREME-MR adapts a joint reconstruction and deformable registration approach for dynamic volumetric MRI reconstruction. Specifically, it reconstructs a reference 3D anatomy $\boldsymbol{I}_{ref}(\boldsymbol{x})$ and solves a cardiorespiratory-resolved dynamic motion field $\boldsymbol{d}(\boldsymbol{x},t)$ with respect to $\boldsymbol{I}_{ref}(\boldsymbol{x})$ to generate dynamic MRIs $\boldsymbol{I}(\boldsymbol{x},t)$. The reference anatomy $\boldsymbol{I}_{ref}(\boldsymbol{x})$ is solved by the spatial implicit neural representation (INR). The dynamic motion fields $\boldsymbol{d}(\boldsymbol{x},t)$ are decomposed into spatial $\boldsymbol{e}_i(\boldsymbol{x})$ and temporal $\boldsymbol{w}_i(t)$ components which are separately estimated by learnable B-spline interpolants and an MLP-based motion encoder, respectively. The motion encoder estimates the time-varying motion coefficients directly from the multi-coil k-space signals extracted from MR signals $\boldsymbol{s}(\boldsymbol{k},t)$ acquired by a 3D golden-mean Koosh-ball trajectory. Therefore, it can be directly applied to the subsequent treatment session for real-time motion estimation, using online time series of MR signals. The dual-task learning is driven by a k-space data consistency loss and regularization losses. MBC: motion basis component.

Figure 1 provides an overview of the DREME-MR workflow and network architecture. DREME-MR consists of a spatial implicit neural representation (INR), learnable B-spline-based interpolants, and a multilayer perceptron (MLP)-based motion encoder, which are responsible for estimating the reference anatomy $\boldsymbol{I}_{ref}(\boldsymbol{x})$, MBCs $\{\boldsymbol{e}_i(\boldsymbol{x})\}_{i=1}^{L}$, and MBC scores $\{\boldsymbol{w}_i(t)\}_{i=1}^{L}$, respectively. During the training stage, the dynamic sequence of volumetric MR images, $\boldsymbol{I}(\boldsymbol{x},t)$, is generated via Eqs. (1-3), using the outputs from the spatial INR, B-spline interpolants, and motion encoder. The dual-task learning is driven by the data consistency and regularization losses in Eq. (4). To calculate the data consistency loss, the estimated k-space data at the sampled k-space points are obtained via NUFFT for comparison with the actual acquisitions. While it may appear that the spatial INR, B-spline interpolants, and motion encoder are separated in DREME-MR, this loss propagates gradients through all components, ensuring end-to-end optimization during training.

The spatial INR was adopted from our previous STINR-MR work (Shao et al. 2024), utilizing MLP-based neural networks with periodic activation functions (i.e., SIREN (Sitzmann et al. 2020)) for implicit neural representation (Mildenhall et al. 2022), by which underlying mappings (e.g., $\boldsymbol{x} \mapsto \boldsymbol{I}_{ref}(\boldsymbol{x})$) are implicitly parametrized by learnable parameters of neural networks. Specifically, the spatial INR takes a voxel coordinate $\boldsymbol{x}$ as input and estimates the MR value of the reference anatomy $\boldsymbol{I}_{ref}(\boldsymbol{x})$ at the queried point $\boldsymbol{x}$. To facilitate learning fine-scale image features, prior to the spatial INR, a learnable hash encoder (Muller et al. 2022) was used to convert the input 3D coordinates $\boldsymbol{x}$ to a feature vector in a high-dimensional feature space. With effective hash encoding, the INR architecture can be made compact and has high learning efficiency to capture a complex anatomy. Default hyper-parameters of the hash encoder (Muller et al. 2022) were used in this work. The spatial INR contains two MLP networks responsible for the real and imaginary components of the MR value, respectively. We found two separate networks attain higher image quality and fewer reconstruction artifacts, compared with a single MLP network with two output channels (Shao et al. 2024). The two networks share the same hash encoder, and each of them has an input, a hidden, and an output layer. The input and hidden layers contain 32 feature channels, and the output layer has a single-channel output. The whole MR volume can be obtained by sequentially querying all voxel coordinates of $\boldsymbol{I}_{ref}(\boldsymbol{x})$.

The MBCs consist of the respiratory $\boldsymbol{e}_i^r(\boldsymbol{x})$ and cardiac $\boldsymbol{e}_i^c(\boldsymbol{x})$ components, parametrized by learnable B-spline-based interpolants, which provide a smooth and sparse representation of the dense MBCs. For $\boldsymbol{e}_i^r(\boldsymbol{x})$, grids of learnable B-spline control points are defined in a multi-resolution scheme. At the *i*-th spatial resolution level, a sparse, uniform grid of control points is set up for each Cartesian component (i.e., *x*, *y*, or *z* direction). Then the MBC $e_{i,k}^r(\boldsymbol{x})$ of the *i*-th level along the *k*-th Cartesian direction at $\boldsymbol{x}$ can be calculated via cubic B-spline interpolation using its neighboring control points. We used three levels of spatial resolutions for respiratory motion. For $\boldsymbol{e}_i^c(\boldsymbol{x})$, a single-level grid of an independent coordinate system enclosing the heart was used for cardiac motion (see Sec. 2.2). The grids of control points for the respiratory MBCs $\boldsymbol{e}_i^r(\boldsymbol{x})$ were 8×8×8, 12×12×12, and 16×16×16, and the grid of control points for the cardiac MBC $\boldsymbol{e}_i^c(\boldsymbol{x})$ was 16×16×16.

The motion encoder is responsible for estimating the temporal components $\boldsymbol{w}_i(t)$ of the motion model in Eq. (2). To fulfill the task of real-time imaging, the motion encoder not only has to solve the time-varying MBC scores for dynamic MRI reconstruction (i.e., learning task 1) but also has to be capable of inferring real-time motion amplitudes reflecting current motion states with low computational latency (i.e., learning task 2), based on online MR signals. Therefore, we designed the motion encoder to estimate the MBC scores directly utilizing acquired multi-coil k-space signals $\boldsymbol{s}(\boldsymbol{k}, t)$ without transforming to the image domain, thus eliminating the use of time-consuming NUFFT operators. In contrast to image-based motion estimation (e.g., (Terpstra et al. 2021, Shao et al. 2022, Wei et al. 2023)), the k-space based approach also avoids the challenge of resolving accurate motion states from severely artifacts-ridden images resulting from extreme undersampling. For phased array acquisition, receiver coils locally probe separate anatomical parts of a subject, thus the multi-coil MR signals offering local motion information of the anatomy. Since every radial spoke of golden-mean radial trajectories passes through the k-space origin, the time series of the zero-frequency components of the MR signals provides a reliable and continuous motion signal. We therefore extracted the zero-frequency components of $\boldsymbol{s}(\boldsymbol{k}, t)$ acquired within each frame and binned the extracted signals into a single bin. Then the binned MR signals of all coils were inputted into the MLP-based motion encoder for MBC score estimation. The motion encoder learns to filter and process the multi-coil zero-frequency signals to estimate the corresponding respiratory and cardiac motion coefficients. The motion encoder consists of 12 MLP networks, each responsible for a Cartesian component of the three-level respiratory MBCs $\boldsymbol{e}_i^r(\boldsymbol{x})$ and cardiac MBC $\boldsymbol{e}_i^c(\boldsymbol{x})$. All networks share the same architecture, comprising three linear layers. Each linear layer, except the last, is followed by a rectified linear unit function. The input layer has twice the number of features as receiver coils, with half for the real, half for the imaginary components of the MR signals. The hidden layers have the same feature number as the input layer, and the output layer has only a single channel, representing the MBC score corresponding to the input signal.

### 3.2 Onboard real-time imaging and motion tracking

After the dual-task learning, DREME-MR yields the current reference anatomy $\boldsymbol{I}_{ref}(\boldsymbol{x})$ and patient-specific motion model, as well as the motion encoder capable of solving time-varying MBC scores based on multi-coil MR signals (i.e., $\boldsymbol{s}(\boldsymbol{k},t) \mapsto \boldsymbol{w}_i(t)$). When deployed during radiation treatment, DREME-MR continuously acquires online MR signals for real-time monitoring, using the same pulse sequence and coil geometry as in the pre-treatment scan. For real-time MR imaging, the reference anatomy is used, together with the real-time DVF $\boldsymbol{d}(\boldsymbol{x},t)$ solved by the cardiorespiratory motion model, to derive the real-time MRI. For real-time target localization, tracking targets in the reference anatomy $\boldsymbol{I}_{ref}(\boldsymbol{x})$ can be contoured, and the target mask replaces the reference anatomy to achieve markerless target localization via DVF-driven propagations.

### 3.3 Regularization loss functions

In addition to the k-space data consistency loss in Eq. (4), several regularization losses were implemented during model training to enable DREME-MR to solve physiologically realistic motion models. The regularization losses include image-domain regularization and motion model regularizations. The image-domain regularization is the total variation (TV) loss that suppresses high-frequency image noise while preserving anatomical edges:

$$L_{TV} = \frac{1}{N_{voxel}} \sum_l \left| \nabla \boldsymbol{I}_{ref}(\boldsymbol{x}_l) \right| , \tag{5}$$

where $N_{voxel}$ is the number of voxels, $l$ is the voxel index, and $\nabla$ denotes the gradient operator which was calculated using forward finite difference.

Three loss functions were introduced to regularize the cardiorespiratory motion model. The first loss function is a normalization loss of MBCs $\boldsymbol{e}_i(\boldsymbol{x})$:

$$L_{MBC} = \frac{1}{3} \sum_{k=x,y,z} \sum_{i=1}^{4} \left( \left| \left\| e_{i,k} \right\|_2^2 - 1 \right|^2 \right) , \tag{6}$$

where the MBC norm $\| \cdot \|_2$ is the L2 norm and calculated analytically using the B-spline control points. By normalizing the MBC norm, this loss function removes the ambiguity of the spatiotemporal decomposition in the low-rank motion model in Eq. (2). The second loss function is the zero-mean loss on the MBC scores $\boldsymbol{w}_i(t)$:

$$L_{ZMS} = \frac{1}{12} \sum_{k=x,y,z} \sum_{i=1}^{4} \left| \frac{1}{N_t} \sum_t w_{i,k}(t) \right|^2 . \tag{7}$$

Essentially, this loss function removes potential time-independent baseline in $\boldsymbol{w}_i(t)$, thereby centering the centroid of the cardiorespiratory motion of the dynamic sequence at the origin of the motion space. Our previous study showed that the zero-mean score loss improves the overall localization accuracy for a digital phantom study (Shao et al. 2025).

The final regularization loss is a temporal frequency constraint on the respiratory $\boldsymbol{w}_i^r(t)$ and cardiac $\boldsymbol{w}_i^c(t)$ MBC scores to promote the decoupling of the two motions in the frequency domain. We found that even with the decoupling of the global respiratory and local cardiac coordinate systems (see Sec. 2.2), the two motions remain entangled in the MBC scores. Therefore, based on the distinct frequencies between respiratory and cardiac motions, we introduced two frequency-domain loss functions that penalize the cardiac and respiratory signals in the respiratory and cardiac MBCs, respectively. The cardiac and respiratory frequency ranges were determined as follows. Prior to model training, the breathing and heartbeat frequencies were identified from the zero-frequency components extracted from the pre-treatment MR scan via Fourier analysis. Next, the frequency bins of the fundamental and high-order harmonics were selected for both motions. Then, during model training, the frequency bins were used to select the cardiac and respiratory frequency components in the estimated respiratory and cardiac MBC scores, and L2 loss functions were applied to suppress these undesired, *cross-over* frequency components. Since the three-level respiratory scores $\boldsymbol{w}_i^r(t)$ vary widely in amplitude depending on motion patterns, resolution levels, and directions, the cardiac signals are defined with respect to baselines to compensate such differences. The cardiac frequency loss function $L_c$ is defined from the respiratory MBC scores as

$$L_c = \frac{1}{N_c} \sum_{\omega \in \nu_c, \omega' \in \nu_b} \sum_{i,k} \left| \mathcal{F}\left[w_{i,k}^r(t)\right](\omega) - \mathcal{F}\left[w_{i,k}^r(t)\right](\omega') \right|^2, \tag{8}$$

where $N_c$ is the sum of all cardiac frequency bins $\nu_c$ of all resolution levels and Cartesian components, $\nu_b$ is the frequency bins for quantifying baselines, and $\mathcal{F}$ denotes the Fourier transform. By minimizing the loss function, it gradually removes the cardiac motion frequencies and signals from the respiratory motion scores. Similarly, the respiratory frequency loss function $L_r$ is defined in the same manner, but without the baseline subtraction from the cardiac MBC scores:

$$L_r = \frac{1}{N_r} \sum_{\omega \in \nu_r} \sum_{i,k} \left| \mathcal{F}\left[w_{i,k}^c(t)\right](\omega) \right|^2, \tag{9}$$

where $N_r$ is the sum of all respiratory frequency bins $\nu_r$ of all Cartesian components.

The total regularization loss function $R$ in Eq. (4) is a weighted sum of the above regularization loss functions:

$$R = \lambda_{TV} L_{TV} + \lambda_{MBC} L_{MBC} + \lambda_{ZMS} L_{ZMS} + \lambda_c L_c + \lambda_r L_r, \tag{10}$$

where λ values denote the weighting factors of the regularization losses empirically determined using the digital phantom simulation study (Sec. 3.5.1). We note that the above regularizations are sufficient for the current framework. However, if overfitting becomes a concern, additional regularizations, such as motion field invertibility, cycle consistency, and volume preservation, can be readily incorporated into the DREME-MR framework as loss terms.

### 3.4 Progressive training scheme and other implementation details

A three-stage, progressive training scheme (Zhang et al. 2023, Shao et al. 2024) was adopted to facilitate dual-task learning and avoid local optima while solving a dynamic sequence of MRIs, as a proper initialization of model components (i.e., the spatial INR and motion model) was found to speed up model training and improve model performance. The progressive training scheme separately warm-starts the spatial INR and the motion model in the first two stages (Stages I and II), followed by a joint training stage (Stage III) of all model components to improve accuracy, consistency, and coherence between the reference anatomy and the motion model. The warm start of the spatial INR at Stage I is further divided into two steps. In the first step of Stage I, an approximate anatomy $\boldsymbol{I}_{aprx}(\boldsymbol{x})$, reconstructed via NUFFT using coil-compressed k-space data of all radial spokes, serves as the training label. The loss function $L_1$ for this step is defined in the image domain:

$$L_1 = \frac{1}{N_{voxel}} \sum_{\boldsymbol{x}} \left\| \boldsymbol{I}_{ref}(\boldsymbol{x}) - \boldsymbol{I}_{aprx}(\boldsymbol{x}) \right\|_1 . \tag{11}$$

The multi-coil k-space data were compressed such that the resulting single-coil data have homogeneous coil sensitivity (Huttinga et al. 2021). Since $\boldsymbol{I}_{aprx}(\boldsymbol{x})$ contains image artifacts resulting from coil compression, anatomical motion, and k-space undersampling, in the second step of Stage I, the similarity loss is changed to the k-space data consistency loss based on all the k-space data (without considering motion), together with the TV regularization loss as defined in Eq. (5):

$$L_2 = \mathbb{E}\left[ \left\| F\left[\boldsymbol{I}_{ref}(\boldsymbol{x})\right] - \boldsymbol{s}(\boldsymbol{k}) \right\|_1 \right] + \lambda_{TV} L_{TV} . \tag{12}$$

This step mitigates the coil-compression and undersampling artifacts from the first step of Stage I, and the remaining artifacts are mainly due to anatomical motion.

After initializing the spatial INR, the motion model is progressively initialized in a multi-resolution manner in Stage II, with the spatial INR and its hash encoder being temporarily frozen. The initialization begins at the lowest spatial resolution $L = 1$ of the respiratory MBCs $\boldsymbol{e}_i^r(\boldsymbol{x})$ and their scores $\boldsymbol{w}_i^r(t)$, and progressively adds the higher spatial resolutions. Each resolution level is initialized over 200 epochs. During the first 50 epochs of each level, the spatial INR and the hash encoder are kept frozen to allow the motion model regularization losses $L_{MBC}$ and $L_{ZMS}$ to stabilize. The same initialization process is then repeated for the next MBC levels. The cardiac components of the motion model are added lastly, after the respiratory components. The initialization of the cardiac MBCs lasts for 50 epochs, with the spatial INR and hash encoder frozen. In total, this stage consists of 850 epochs. The loss functions at this stage includes the k-space data consistency loss in Eq. (4) and all the regularization losses (i.e., the image TV loss $L_{TV}$, the motion model losses $L_{MBC}$ and $L_{ZMS}$, and the frequency guidance losses $L_c$ and $L_r$) in Eq. (10). For Stage III, all components of DREME-MR are activated for joint learning, as illustrated in Figure 1. Similar to Stage II, the loss functions at this stage are defined in Eqs. (4) and (10). This stage allows the motion model to refine its representations by simultaneously optimizing the spatial INR, hash encoder, and motion components, leading to improved overall performance.

Other implementation details of our algorithm are in order: (a) The DREME-MR framework was implemented using the PyTorch library v1.13 (Paszke et al. 2019), and the NUFFT operator was adopted from the TorchKbNufft library (Muckley et al. 2020). Adam optimizer was used for batched model training. (b) MRI k-space signals $\boldsymbol{s}(\boldsymbol{k}, t)$ were sequenced into MRI frames, and the frames were randomly sampled for each training epoch when k-space losses are computed, with a batch size of 32. (c) No early stopping criteria were used. Instead, model training proceeded for a fixed number of iterations for each stage, empirically determined based on the XCAT simulation study. The numbers of epochs for the first (Eq. 11) and second (Eq. 12) steps of the first stage were 500 and 1,300, respectively. The numbers of epochs for Stages II and III were 850 and 3,650, respectively. (d) Since the raw k-space signals $\boldsymbol{s}(\boldsymbol{k}, t)$ of different coils involve wide ranges of variations, z-score normalization was applied to the real- and imaginary-channel of $\boldsymbol{s}(\boldsymbol{k}, t)$ prior to feeding them into the motion encoder. (e) The learning rate was fixed throughout each stage of the progressive training. Due to the change of losses (Eqs. (11) and (12)) during the warm start, the learning rates of the spatial INR were reset at the second step of Stage I. We used learning rates of $2 \times 10^{-4}$ for the first step of Stage I, and $5 \times 10^{-5}$ for the second step of Stage I (and the following Stages II and III), respectively. For the motion model (i.e., the B-spline interpolants and MLP-based motion encoder), we used a learning rate of $5 \times 10^{-4}$. (f) The weighting factors in Eq. (10) were determined empirically using the digital phantom study in Sec. 2.6.1, the numerical values were $\lambda_{TV} = 2 \times 10^{-6}$, $\lambda_{MBC} = 1 \times 10^{-2}$, $\lambda_{ZMS} = 1 \times 10^{-4}$, $\lambda_c = 1 \times 10^{-1}$, and $\lambda_r = 5 \times 10^{-2}$. (g) Since a localized cardiac coordinate system was introduced in our motion model (see Sec. 2.1), we determined the dimension of the cardiac coordinate system by empirical searching. We used a box of 54 × 48 × 54 voxels for the digital phantom (Sec. 2.6.1) and a box of 60 × 50 × 48 voxels for the human subject study (Sec. 2.6.2). (h) The cardiac coordinate system may cause discontinuities on the border of cardiac MBCs (and thus the resultant DVFs after combination with the respiratory motion), as the control points on the edges are free learnable parameters. In this work, the continuity of DVFs was achieved by enforcing the values of the control points at the boundaries of the cardiac coordinate system to zero such that $\boldsymbol{e}_i^c(\boldsymbol{x})$ is also zero there.

## 3.5 Evaluation datasets and schemes

DREME-MR was evaluated by a digital phantom-based simulation study and a human subject study. The simulation study used the extended cardiac torso (XCAT) digital phantom (Segars et al. 2010) which provides 'ground-truth' images for algorithm design, hyper-parameter tuning, and model validation. After the validation, we further tested DREME-MR on a healthy human subject to assess its potential for clinical adoption. We separately discuss the details and preprocessing steps of both studies below.

### *3.5.1 XCAT simulation study*

To assess the capability of DREME-MR to capture different types of irregularity in respiratory motion in a 'one-shot' learning manner, we simulated six motion scenarios with various types of irregular motion variations in breathing frequency, amplitude, and baseline (Table 1). To add complexity and prevent potential data leakage, each motion scenario uses different combinations of superior-anterior (SI) motion amplitudes (18-24 mm) and anterior-posterior (AP) motion amplitudes (10-12 mm). In order to evaluate the tracking accuracy for respiratory motion, a lung tumor with a 30-mm diameter was inserted into the lower lobe of the right lung and served as the tracking target. For cardiac motion, the default XCAT heart motion curve with a 1-sec period was used for all scenarios (X1-X6), as cardiac motion is generally more consistent and stable than respiratory motion. The XCAT volumes contain 150 × 150 × 150 voxels with a $3 \times 3 \times 3\ \text{mm}^3$ resolution, covering the thoracic-abdominal region. The image intensities were normalized to the range [0, 1]. Since the XCAT phantom only renders real-valued MR volumes, phased angles were randomly assigned to organs and tissues to simulate complex signals. In total, 1,860 frames of dynamic MRI volumes were generated for each motion scenario assuming a 3-min MR scan, corresponding to a temporal resolution of 96.8 ms.

**Table 1**. Motion parameters and characteristics of the six types of motion scenarios (X1-X6) in the XCAT simulation study.

| Motion scenario | Superior-inferior motion amplitude (mm) | Anterior-posterior motion amplitude (mm) | Motion characteristics |
|---|---|---|---|
| X1 | 20 | 10 | Regular breathing with small amplitude variations |
| X2 | 24 | 12 | Sudden baseline shift in the middle of MR scan |
| X3 | 18 | 10 | Amplitude variations with slow baseline drift |
| X4 | 20 | 12 | Decreasing breathing frequency with increasing amplitude |
| X5 | 22 | 10 | Slow breathing with amplitude variations |
| X6 | 23 | 11 | Combinations of baseline drift, amplitude and frequency variations |

After generating the 'ground-truth' XCAT dynamic MRI volumes, we simulated the corresponding k-space data from these MRI volumes. The pulse sequence was a steady-state spoiled gradient echo sequence with a TR of 4.4 ms. From each sequential dynamic MRI volume (96.8 ms temporal resolution), 22 spokes of k-space data were simulated. The k-space trajectory followed a 3D golden-mean Koosh-ball radial pattern with 150 readout points along each radial spoke. We simulated 24 coils arranged in three rows stacked along the SI direction. For each row, eight coils were concentrically distributed 300-mm from the longitudinal axis of the XCAT phantom. The middle row aligns with the center of the XCAT volumes, and the top and bottom rows were shifted in the superior and inferior directions by 90 mm, respectively. Each coil had a square shape with a 180-mm length. The sensitivity maps of the coils were calculated using the Biot-Savart law under the quasi-static limit (Roemer et al. 1990, Wang et al. 1995). The undersampling ratio is about 2,500 estimated based on the assumption of uniform angular sampling in the radial and azimuthal directions.

The outcomes of the two learning tasks were separately evaluated. For learning task 1, DREME-MR was separately trained on the six motion scenarios (X1-X6), and the image quality of the reconstructed dynamic MRI and the accuracy of the derived motion were evaluated on the same motion scenarios. To properly assess learning task 2, the DREME-MR model trained on a motion scenario (e.g., X1) was crossly tested on the other scenarios (i.e., X2-X6) to demonstrate its generalizability to unseen motion scenarios. The image quality was evaluated by intensity-based metrics: relative error (RE), contrast error (ER), and structural similarity index measure (SSIM) (Wang et al. 2004). The relative error is defined as mean relative intensity difference between the prediction and ground-truth images:

$$RE = \frac{1}{N_t}\sum_t \sqrt{\frac{\sum_x \left(|\boldsymbol{I}(\boldsymbol{x},t)| - |\boldsymbol{I}_{gt}(\boldsymbol{x},t)|\right)^2}{\sum_x |\boldsymbol{I}_{gt}(\boldsymbol{x},t)|^2}}, \quad (13)$$

where $N_t$ is the number of frames in the dynamic sequence, and $\boldsymbol{I}_{gt}(\boldsymbol{x}, t)$ is the 'ground-truth' volumes. The contrast error is defined as

$$CE = 1 - \frac{1}{N_t}\sum_t \frac{2\sigma(t)\sigma_{gt}(t) + \epsilon}{\sigma^2(t) + \sigma_{gt}^2(t) + \epsilon}, \quad (14)$$

where $\sigma(t)$ and $\sigma_{gt}(t)$ respectively are the predicted and ground-truth local intensity standard deviations at frame $t$, and $\epsilon = 9.18 \times 10^{-4}$ is a parameter stabilizing the division. The local intensity standard deviation was calculated by a window size of 11×11×11. This metric quantifies the contrast restoration performance with $CE = 0$ indicating a perfect restoration. SSIM is an intensity-based metric commonly used for quantifying perceived image quality by combining luminance, contrast, and structure.

To evaluate motion tracking accuracy, we considered the lung tumor as the tracking target for respiratory motion and the left ventricle (LV) as the tracking target for cardiac motion. The tracking accuracy was evaluated by contour-based metrics, including target center-of-mass error (COME), Dice similarity coefficient (DSC), and 95-percentile Hausdorff distance (HD95). The COME is defined as the center-of-mass difference between the estimated and 'ground-truth' target centers-of-mass. DSC quantifies the overlap of the estimated and 'ground-truth' contours. HD95 quantifies the estimated and 'ground-truth' target surface distance. We contoured the tracking targets from the reconstructed reference anatomy $\boldsymbol{I}_{ref}(\boldsymbol{x})$ of each motion scenario, propagated the tracking masks by the estimated DVFs $\boldsymbol{d}(\boldsymbol{x})$, and compared the propagated masks on each dynamic volume with the 'ground-truth' counterparts. In addition to motion tracking accuracy, the biomechanical plausibility of the DVFs is assessed using two Jacobian-based metrics. The first metric is the standard deviation of the logarithm of the Jacobian determinant (SD*log*(J)), which reflects the smoothness and incompressibility of motion fields (Hering et al. 2023). As most biological tissues are nearly incompressible during anatomical motion, smaller SD*log*(J) values suggest better preservation of local tissue volume. The second metric is the percentage of the negative Jacobian determinant (J<0), which indicates local folding or inversion of tissue. Both metrics were evaluated over the entire anatomy, excluding the lungs, which are compressible during respiration.

In addition to the motion-tracking study at fixed temporal resolution, we also investigated the DREME-MR-induced tracking latency as a function of temporal resolution. We simulated the same XCAT motion scenarios (X1-X6) with a 48.4-ms temporal resolution. DREME-MR was then trained and tested at three temporal resolutions: 48.4 ms, 96.8 ms, and 145.2 ms by grouping the 48.4-ms k-space acquisition into low-resolution frames. This grouping introduces intra-frame k-space inconsistency and motion artifacts, which may lead to latency and/or instability in motion tracking. To quantify these effects, we evaluated tumor and LV tracking accuracy and compared the DREME-MR-predicted trajectories with the 'ground-truth' trajectories. The temporal latency was evaluated using the Pearson correlation coefficients.

Considering that k-space acquisition is inherently noisy in a clinical setting, we performed an additional XCAT study to evaluate the robustness of DREME-MR to noise. Gaussian noise was introduced in two stages to simulate two noise sources: body thermal noise and receiver coil electronic noise. First, complex-valued Gaussian noise with a standard deviation of 0.1 was added to the image-domain dynamic XCAT volumes to simulate thermal noise. This noise level resulted in a mean peak signal-to-noise ratio of 26.0 dB across the motion scenarios. The corresponding noise-corrupted multi-coil k-space data were then computed, and a second stage of complex-valued Gaussian noise with a standard deviation of 0.01 was added to simulate electronic noise in the acquisition system. To ensure that each motion scenario has a different noise realization, a unique random seed was used for each scenario. Motion tracking accuracy was evaluated and compared with noise-free results. Due to space limitations, detailed results of the latency and noise studies are presented in the Supplementary Materials.

### *3.5.2 Human subject study*

The human dataset contains a free-breathing MR scan of a healthy subject covering the thoracic-abdominal region from the University Medical Center Utrecht (Huttinga et al. 2021). The data were acquired by a 1.5-T MRI scanner (Ingenia, Philips Healthcare) and are openly accessible. The pulse sequence and k-space trajectory used were the same as those in the XCAT study. The repetition and echo times were 4.4 ms and 1.8 ms, respectively. The total scan time was 297.4 s, resulting in 67,280 radial spokes, each with 232 readout points. The first 900 spokes were discarded to allow the scanner to reach a steady state. The k-space signals were measured by 24 receiver coils, with 12 anterior and 12 posterior coils. The dataset includes the complex-valued k-space data, k-space trajectory, coil sensitivity maps, and noise covariance matrix.

Compared with the XCAT motion-tracking study, the k-space data contain a high level of noise. Therefore, in contrast to the pre-defined k-space spoke grouping (22 spokes per MRI volume) as used in the XCAT study, we adopted on-the-fly k-space grouping during model training to improve model robustness to noise. During the model training, an MRI frame was defined as 34 consecutive spokes (=149.6 ms) which were randomly grouped from the whole sequence for reconstruction, and 32 such frames were extracted for each training batch. The reconstruction volume had 150 × 150 × 150 voxels with 3.0 × 3.0 × 3.0 $mm^3$ resolution.

Since the dataset does not include an independent onboard MR scan for real-time motion monitoring evaluation, we partition the k-space data into a training and a testing set to evaluate the reconstruction and real-time imaging accuracy. The training set includes the first 75% of k-space data, while the remaining 25% were reserved for real-time tracking evaluation. As no 'ground-truth' images were available for the human study, we visually inspected the reconstructed dynamic MR images. For quantitatively evaluating motion tracking, we calculated the liver and heart LV centers-of-mass trajectory and compared them with motion surrogate signals directly extracted from the k-space data. The surrogate signals for the cardiac and respiratory motions were separately extracted. First, the zero-frequency components of all coils were extracted from the k-space data, and then every 34 consecutive spokes were grouped into frames. Next, the respiratory and cardiac surrogate signals of each coil were extracted by applying low-pass and high-pass filters to the binned signals, using a cutoff frequency of 0.8 Hz. (The frequencies of the respiratory and cardiac motions were 0.26 Hz and 1.4 Hz, respectively, well separated from the 0.8-Hz cutoff frequency.) Finally, the filtered signals which have the highest Pearson correlation coefficients with the liver or LV motion trajectories solved by DREME-MR (with 100% k-space data) were selected as the surrogate signals.

### 3.6 Comparison and motion model studies

Dynamic volumetric MRI reconstruction and real-time imaging remain an active research area, and currently, no 'gold-standard' methods are available in clinics. To the best of our knowledge, no other 'one-shot' dynamic or real-time volumetric MR reconstruction studies have been reported that can simultaneously resolve cardiac and respiratory motion. Accordingly, we compared DREME-MR with principal component analysis (PCA)-based methods, which also enable real-time imaging, as well as two additional dynamic MRI reconstruction methods in the XCAT simulation study.

The PCA-based method constructs patient-specific motion models by applying PCA to intra-phase DVFs derived from 4D-MRI. Consequently, the PCA-based method has two variants, called online PCA and offline PCA, depending on the sources of the 4D-MRI. Online PCA reconstructs 4D-MRIs using pre-treatment scans, whereas offline PCA assumes the availability of separate, artifact-free 4D-MRIs. For online PCA, due to the limited k-space data of the pre-treatment scans (~4 min), sorting the k-space data into both cardiac and respiratory motion bins for 4D-MRI reconstruction will lead to severe undersampling artifacts. Thus, we disregarded cardiac motion and sorted the k-space data into 10 respiratory phases to ensure sufficient sampling in each respiratory phase for reconstruction. As a result, online PCA is not expected to resolve cardiac motion. For offline PCA, we simulated a 4D-MRI of XCAT to resolve both cardiac and respiratory motion, assuming that the scan time is sufficient to allow adequate sampling during the treatment simulation stage. Thus, the ability of the offline PCA-based method to capture cardiac and respiratory motion is evaluated. Specifically, offline PCA models both respiratory and cardiac motions, with each divided into 10 phases. To avoid excessively enumerated cardiorespiratory phases (i.e., 10×10 phases) in the PCA-based motion model, we adopted the same sequential registration approach used in DREME-MR, where cardiac deformable registration was first performed, followed by respiratory deformable registration. In detail, the cardiac PCA model was derived from 10 cardiac phases at the end-of-exhale respiratory phase, using the diastolic cardiac phase as the reference phase. Then, the respiratory PCA model was derived using 10 respiratory phases with the same cardiac phase (i.e., the diastole phase), using the end-of-exhale phase as the reference phase. These two PCA models allow sequential registration to reconstruct cardiorespiratory motion-resolved images. Through this strategy, only 9+9=18 registrations are needed, compared to 99 registrations if the sequential registration framework were not employed. The motion coefficients in the PCA-based motion model were estimated using an MLP-based motion encoder similar to that employed in DREME-MR. We note that online PCA and offline PCA methods are similar to our previous dynamic MRI reconstruction framework STINR-MR (Shao et al. 2024), with modifications that enable cardiac motion tracking (for offline PCA) and real-time imaging via the MLP-based motion encoder.

In addition to the PCA-based models, DREME-MR was compared with two other dynamic reconstruction methods. The first method is Extreme MRI (Ong et al. 2020), which uses multiscale low-rank matrix factorization to represent dynamics at multiple scales. Since Extreme MRI is not a registration-based approach, its motion tracking accuracy was evaluated by thresholding the lung tumor in each reconstructed frame. Due to the complexity of cardiac anatomy, thresholding was unable to accurately delineate LV, and therefore, motion tracking accuracy for cardiac motion was not assessed. The second method is MR-MOTUS (Huttinga et al. 2020). Since the current MR-MOTUS method is unable to resolve cardiac motion, only lung tumor localization accuracy was evaluated.

In addition to the above comparison study, we perform a study of different variants of cardiorespiratory motion models within the DREME-MR framework. The cardiorespiratory motion model in Sec. 2.2 decouples the cardiac coordinate system from the respiratory coordinate system, and the deformable registration is performed in the order of the cardiac deformable registration followed by the respiratory deformable registration (Eq. (3)). However, there is another equivalent motion model where the two registrations are performed in the opposite order. Because theoretically, the two motion models are equivalent, it is unclear which approach is more favorable. We therefore trained a variant of DREME-MR, based on the opposite-order registration. The variant of DREME-MR is called DREME-MR$_{\mathrm{R1C2}}$, where the subscript "R1C2" indicates the order of the sequential registration. We also considered another variant of the motion model where the respiratory $\boldsymbol{d}_r$ and cardiac $\boldsymbol{d}_c$ DVFs are additive (i.e., $\boldsymbol{d}(\boldsymbol{x},t) = \boldsymbol{d}_r(\boldsymbol{x},t) + \boldsymbol{d}_c(\boldsymbol{x},t)$). This motion model can be viewed as adding another level of MBC that is specialized for the heart to the multi-resolution respiratory MBCs. This variant is called DREME-MR$_{\mathrm{R+C}}$, where the subscript "R+C" indicates that the respiratory and cardiac motions are summed. For the above comparison studies, non-parametric Wilcoxon signed-rank tests between DREME-MR and the other models were performed to evaluate the significance levels of observed differences in image quality and motion tracking accuracy.

## 4. Results

### 4.1 The XCAT simulation study

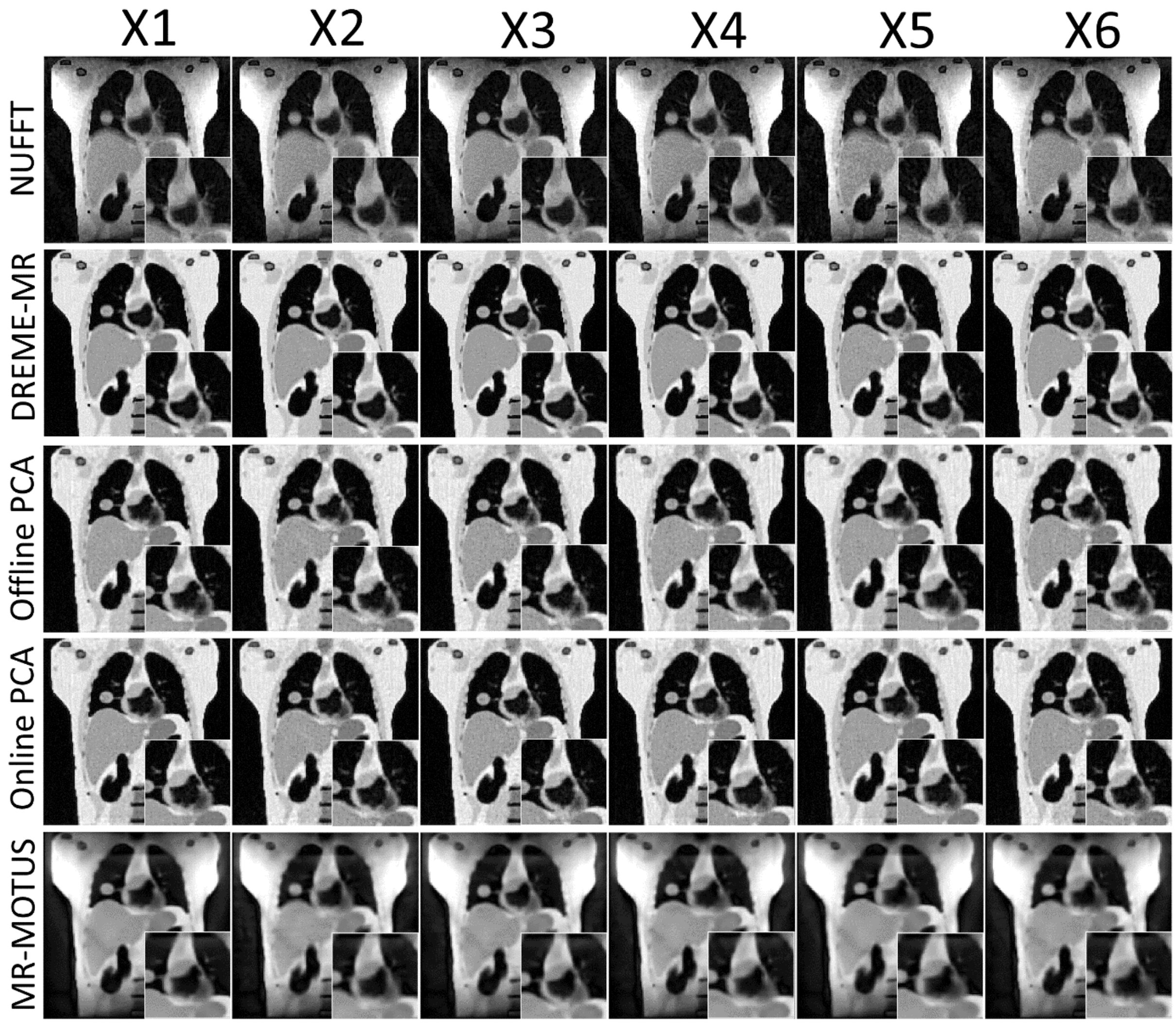

**Figure 2.** Comparison of reference anatomies reconstructed using NUFFT, DREME-MR, PCA-based motion models, and MR-MOTUS, for the six motion scenarios (X1-X6). Zoomed-in views highlight the heart region for detailed comparison. Prominent shading artifacts in the peripheral regions of the NUFFT and MR-MOTUS reconstruction are caused by inhomogeneous coil sensitivity and coil compression.

Figure 2 presents the reconstructed reference anatomies for the six motion scenarios (X1-X6) in the XCAT study. The first row shows NUFFT-based reconstructions, using all coil-compressed k-space data without motion correction, which exhibit significant shading artifacts due to inhomogeneous coil sensitivity maps and image blurriness due to the cardiorespiratory motion. The second row presents the DREME-MR results, where these artifacts and motion-induced blurring are substantially reduced, leading to better-defined anatomical structures. Note that, because the data-driven motion model does not control/designate the motion state of the reference anatomy during model training, no 'ground-truth' reference anatomy is available for comparison. PCA-based methods exhibit slightly noisier reconstructions than DREME-MR. The reference anatomies of MR-MOTUS also exhibit shading artifacts in the peripheral regions, because MR-MOTUS employs coil-compressed data for dynamic reconstruction. The anatomies have less noise but are slightly over-smoothed because of the wavelet-based regularization. Since Extreme MRI is not a registration-based reconstruction, there are no reference anatomies to present.

Table 2 summarizes the results of image quality evaluation for both dynamic reconstruction and real-time imaging tasks, averaged over all training and testing scenarios. Essentially all variants of DREME-MR achieved comparable performance on image quality. We found Extreme MRI exhibits flickering temporal artifacts (Ong et al. 2020), as no mechanism is implemented to maintain consistent image contrast across frames. MR-MOTUS has the worst image quality, partially due to the significant shading artifacts and over-smoothed reconstruction. All Wilcoxon signed-rank tests between DREME-MR and other methods yielded p-values < $10^{-3}$.

**Table 2.** Image reconstruction quality evaluation of the XCAT study. The tasks of dynamic MRI reconstruction and real-time imaging are evaluated separately. Image quality is assessed using intensity-based metrics: relative error (RE), contrast error (CE), and structural similarity index measure (SSIM). For Extreme MRI and MR-MOTUS, only the dynamic reconstruction task is evaluated, as these methods do not support real-time imaging. The results are reported as mean±SD. Bold indicates the best model(s). The arrows are pointing in the direction of improved accuracy.

| Task | Metric | Motion model study | | | Comparison study | | | |
|---|---|---|---|---|---|---|---|---|
| | | DREME-MR$_{R+C}$ | DREME-MR$_{R1C2}$ | DREME-MR | Offline PCA | Online PCA | Extreme MRI | MR-MOTUS |
| Dynamic reconstruction | RE↓ | **0.162±0.010** | **0.162±0.010** | **0.162±0.010** | 0.195±0.011 | 0.187±0.009 | 0.201±0.025 | 0.424±0.036 |
| | CE↓ | **0.025±0.004** | 0.026±0.004 | **0.025±0.004** | 0.046±0.003 | 0.041±0.003 | 0.034±0.006 | 0.149±0.011 |
| | SSIM↑ | **0.852±0.012** | **0.852±0.012** | **0.852±0.012** | 0.801±0.013 | 0.817±0.011 | 0.808±0.034 | 0.509±0.076 |
| Real-time imaging | RE↓ | **0.164±0.009** | **0.164±0.009** | **0.164±0.009** | 0.198±0.011 | 0.189±0.009 | N/A | N/A |
| | CE↓ | **0.025±0.004** | **0.025±0.004** | **0.025±0.004** | 0.046±0.003 | 0.041±0.003 | N/A | N/A |
| | SSIM↑ | **0.850±0.011** | **0.850±0.012** | **0.850±0.012** | 0.799±0.013 | 0.815±0.011 | N/A | N/A |

Table 3 summarizes the motion tracking accuracy for respiratory and cardiac motions. The p-values for all Wilcoxon signed-rank tests between DREME-MR and other variants were <$10^{-3}$. Overall, DREME-MR outperformed all other motion models in both reconstruction and real-time imaging tasks. For the reconstruction task, all DREME-MR variants achieved sub-voxel tumor tracking accuracy (~0.6 mm). However, the LV COMEs (~1.3 mm) were worse than the tumor tracking, likely due to the complexity of the heart motion that involves both respiration and heartbeat. For DSC and HD95, all variants achieved similar scores. Offline PCA achieved a good tumor localization accuracy (~0.9 mm) but showed a large decrease in LV tracking accuracy (~3.9 mm) for the dynamic reconstruction task, indicating that the sequential registration approach (Sec. 3.6) is ineffective for PCA-based cardiorespiratory motion model. Online PCA exhibited reduced tumor tracking accuracy in both dynamic reconstruction and real-time imaging tasks (~1.4 mm), when compared with offline PCA. We found baseline shifts in the online PCA-solved motion trajectories, which are likely caused by motion sorting and undersampling artifacts. Extreme MRI had the worst tumor tracking performance. We found that its multiscale low-rank factorization is effective to capture coarse-scale image contrast variation (e.g., liver/bowel motion), but it fails to resolve variations of small anatomic features (e.g., the lung tumor), thus resulting in substantial localization error. MR-MOTUS achieved moderate tracking accuracy (~2.4 mm). Since MR-MOTUS decouples reference anatomy reconstruction from dynamic MRI reconstruction, the reference anatomy is not further refined during the dynamic reconstruction process. As a result, any artifacts present in the reference anatomy can propagate into the motion model, leading to increased errors in the reconstructed dynamics. For the real-time imaging task, a mild decrease in tracking accuracy for both tumor and LV is observed, which is expected due to unseen motion variations (Sec. 3.5.1). Nevertheless, the tumor tracking accuracy still achieved sub-voxel accuracy (~0.7 mm), demonstrating that DREME-MR can effectively estimate motion in previously unseen scenarios. Among the variants, while DREME-MR had slightly lower accuracy for tumor tracking, it outperformed the others in estimating cardiac motion.

Table 4 presents the mean DVF quality for DREME-MR. No difference was observed in the mean SD*log*(J) between the reconstruction and real-time imaging tasks. The percentage of voxels with negative Jacobian determinants showed a slight increase in real-time imaging. Overall, the small SD*log*(J) and J<0 values indicate that DREME-MR produces biomechanically plausible motion fields with negligible unrealistic deformations.

**Table 3.** Summary of motion tracking accuracy in the XCAT study. The tracking targets are the lung tumor and left ventricle (LV) for the respiratory and cardiac motions, respectively. No LV localization accuracy is evaluated for online PCA, Extreme MRI, and MR-MOTUS, as either the related motion model is unable to resolve cardiac motion (online PCA, MR-MOTUS), or the method is not registration-based, rendering it difficult to evaluate motion tracking accuracy for complex anatomies (LV) that are difficult to segment (Extreme MRI). Extreme MRI and MR-MOTUS are dynamic MRI reconstruction algorithms, thus having no real-time imaging capability. The results are reported as mean±SD. Bold indicates the best model(s). The arrows are pointing in the direction of improved accuracy.

| Task | Tracking target and metric | Motion model study | | | Comparison study | | | |
|---|---|---|---|---|---|---|---|---|
| | | DREME-MR$_{R+C}$ | DREME-MR$_{R1C2}$ | DREME-MR | Offline PCA | Online PCA | Extreme MRI | MR-MOTUS |
| Dynamic reconstruction | Tumor COME (mm) ↓ | 0.68±0.31 | 0.65±0.30 | **0.62±0.29** | 0.92±0.45 | 1.35±0.50 | 8.08±8.21 | 2.42±1.75 |
| | Tumor DSC ↑ | **0.93±0.01** | **0.93±0.01** | **0.93±0.01** | 0.92±0.02 | 0.89±0.03 | 0.84±0.08 | 0.87±0.07 |
| | Tumor HD95 (mm) ↓ | 3.00±0.09 | **2.99±0.16** | **2.99±0.20** | **2.99±0.15** | 3.05±0.25 | 7.01±6.51 | 3.60±1.34 |
| | LV COME (mm) ↓ | 1.34±0.79 | 1.29±0.76 | **1.24±0.68** | 3.93±1.51 | N/A | N/A | N/A |
| | LV DSC ↑ | **0.93±0.01** | **0.93±0.01** | **0.93±0.01** | 0.88±0.02 | N/A | N/A | N/A |
| | LV HD95 (mm) ↓ | **3.00±0.02** | **3.00±0.03** | **3.00±0.01** | 4.69±0.96 | N/A | N/A | N/A |
| Real-time imaging | Tumor COME (mm) ↓ | **0.72±0.39** | 0.76±0.40 | 0.73±0.38 | 1.02±0.52 | 1.36±0.52 | N/A | N/A |
| | Tumor DSC ↑ | **0.93±0.01** | **0.93±0.01** | **0.93±0.01** | 0.92±0.02 | 0.89±0.03 | N/A | N/A |
| | Tumor HD95 (mm) ↓ | **2.99±0.16** | **2.99±0.14** | **2.99±0.20** | 3.00±0.13 | 3.06±0.27 | N/A | N/A |
| | LV COME (mm) ↓ | 1.75±1.17 | 1.75±1.20 | **1.69±1.12** | 4.01±1.61 | N/A | N/A | N/A |
| | LV DSC ↑ | **0.92±0.02** | **0.92±0.02** | **0.92±0.02** | 0.88±0.03 | N/A | N/A | N/A |
| | LV HD95 (mm) ↓ | 3.20±0.55 | 3.21±0.57 | **3.19±0.57** | 4.79±1.32 | N/A | N/A | N/A |

**Table 4**. DREME-MR deformation vector field quality evaluation for the XCAT study. The standard deviation of the logarithm of the Jacobin determinant (SD*log*(J)) and the percentage of voxels with negative Jacobian determinants (J<0) are evaluated to assess the smoothness and physiological reality of the motion fields. The results were reported as mean±SD.

| Task | SD*log*(J) | J<0 (%) |
|---|---|---|
| Dynamic reconstruction | 0.037±0.015 | 0.000±0.001 |
| Real-time imaging | 0.037±0.017 | 0.002±0.009 |

Figure 3 compares the center-of-mass trajectories of the lung tumor and LV in the SI and AP directions at three temporal resolutions. The DREME-MR model was trained on the regular breathing scenario (X1) and crossly tested on all scenarios (X1-X6). Different curves represent different temporal resolutions. The results show that DREME-MR can capture various types of motion irregularity with different ranges of motion amplitudes (Table 1), and the motion tracking accuracy remained comparable across all temporal resolutions. Only minor deviations were observed at respiratory peaks, and no observable latency was detected. From Figure 3(b), we can observe that the respiratory motion remains the dominant motion in both SI and AP directions, especially for the SI direction. Table 5 summarizes the Pearson correlation coefficients between the 'ground-truth' and DREME-MR predicted motion trajectories at three temporal resolutions. The consistently high Pearson correlation coefficients across both tasks and all resolutions indicate good temporal alignment between the predicted and 'ground-truth' trajectories.

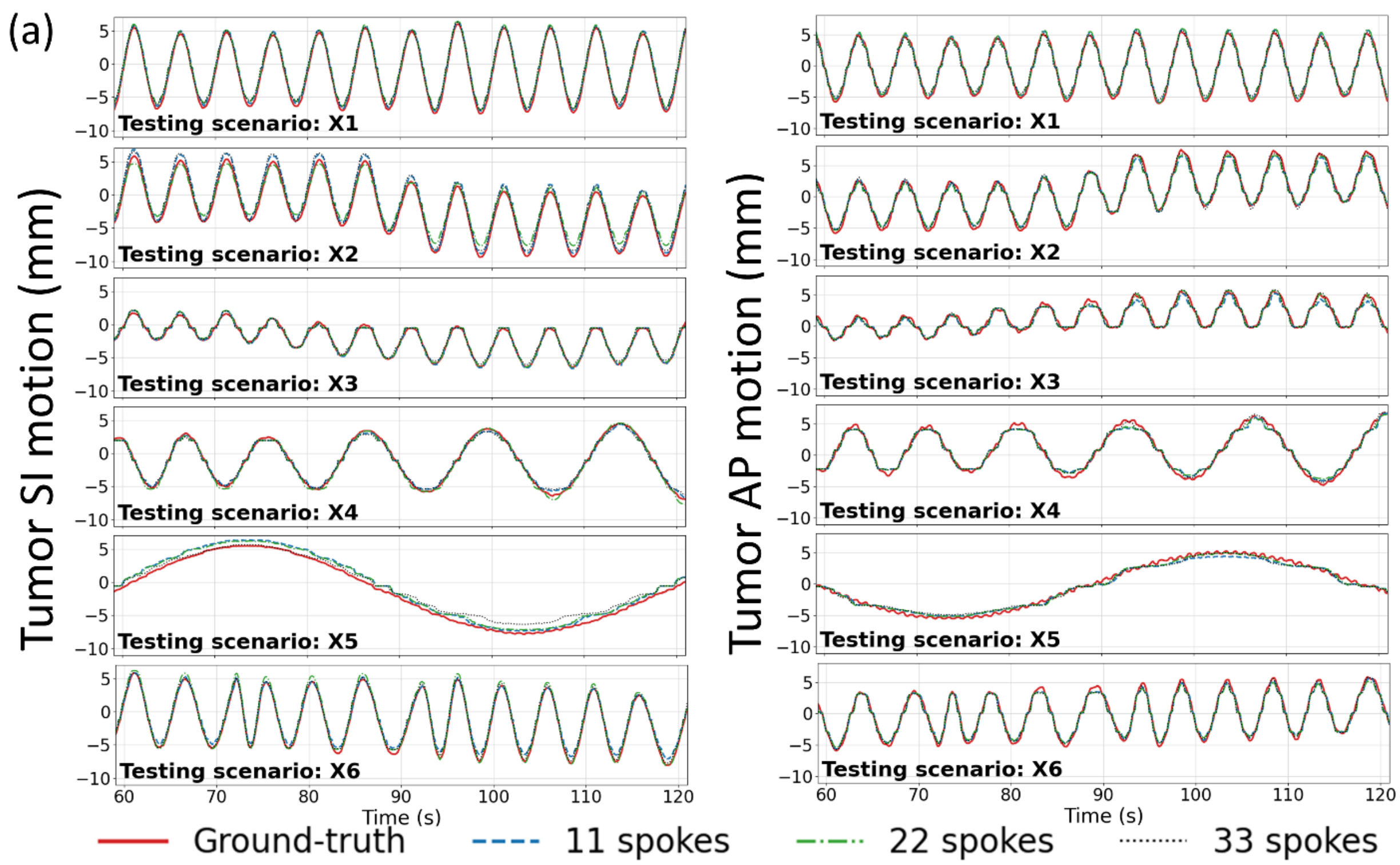

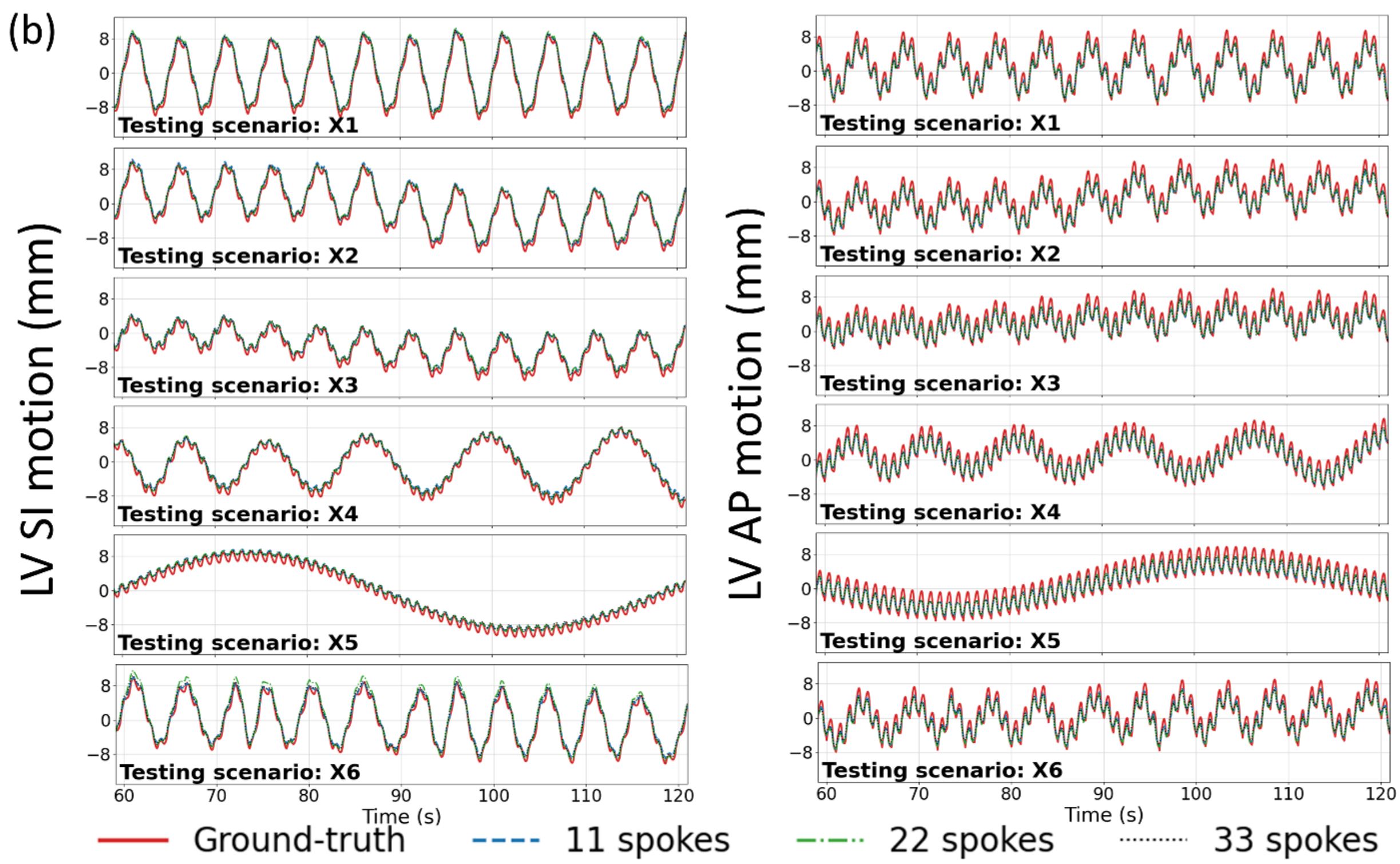

**Figure 3.** Center-of-mass trajectories of (a) lung tumor and (b) heart LV at three temporal resolutions in the XCAT study. The DREME-MR model was trained on the X1 scenario and tested across all scenarios (X1-X6). The first rows of (a) and (b) present the solved motion trajectories for the reconstruction task, and the other rows present the estimated trajectories for the real-time imaging task. Trajectories based on 11, 22, and 33 spokes correspond to temporal resolutions of 48.4 ms, 96.8 ms, and 145.2 ms, respectively. Due to space constraints, only results corresponding to frame indices 500 to 1300 are displayed.

**Table 5**. Pearson correlation coefficients of the system latency study.

| Task | 11 spokes | | 22 spokes | | 33 spokes | |
|---|---|---|---|---|---|---|
| | AP | SI | AP | SI | AP | SI |
| Dynamic reconstruction | 0.993±0.004 | 0.998±0.002 | 0.992±0.003 | 0.997±0.002 | 0.993±0.002 | 0.998±0.002 |
| Real-time imaging | 0.992±0.006 | 0.998±0.001 | 0.991±0.006 | 0.997±0.002 | 0.992±0.006 | 0.998±0.001 |

Figure 4 presents representative cases of (a-b) dynamic MRI reconstruction and (c-d) real-time imaging, with zoomed-in coronal views highlighting the cardiac region. In both tasks, limited discrepancies are observed in the bowel region, likely due to its complex anatomy and deformation. Overall, dynamic reconstruction (a-b) yields slightly better agreement with the ‘ground truth’. Real-time imaging (c-d) exhibits increased reconstruction differences, especially for the X2 testing scenario, which involves large motion amplitudes and baseline shifts. These differences are more pronounced in the diaphragm and the LV regions, as both structures are heavily impacted by motion.

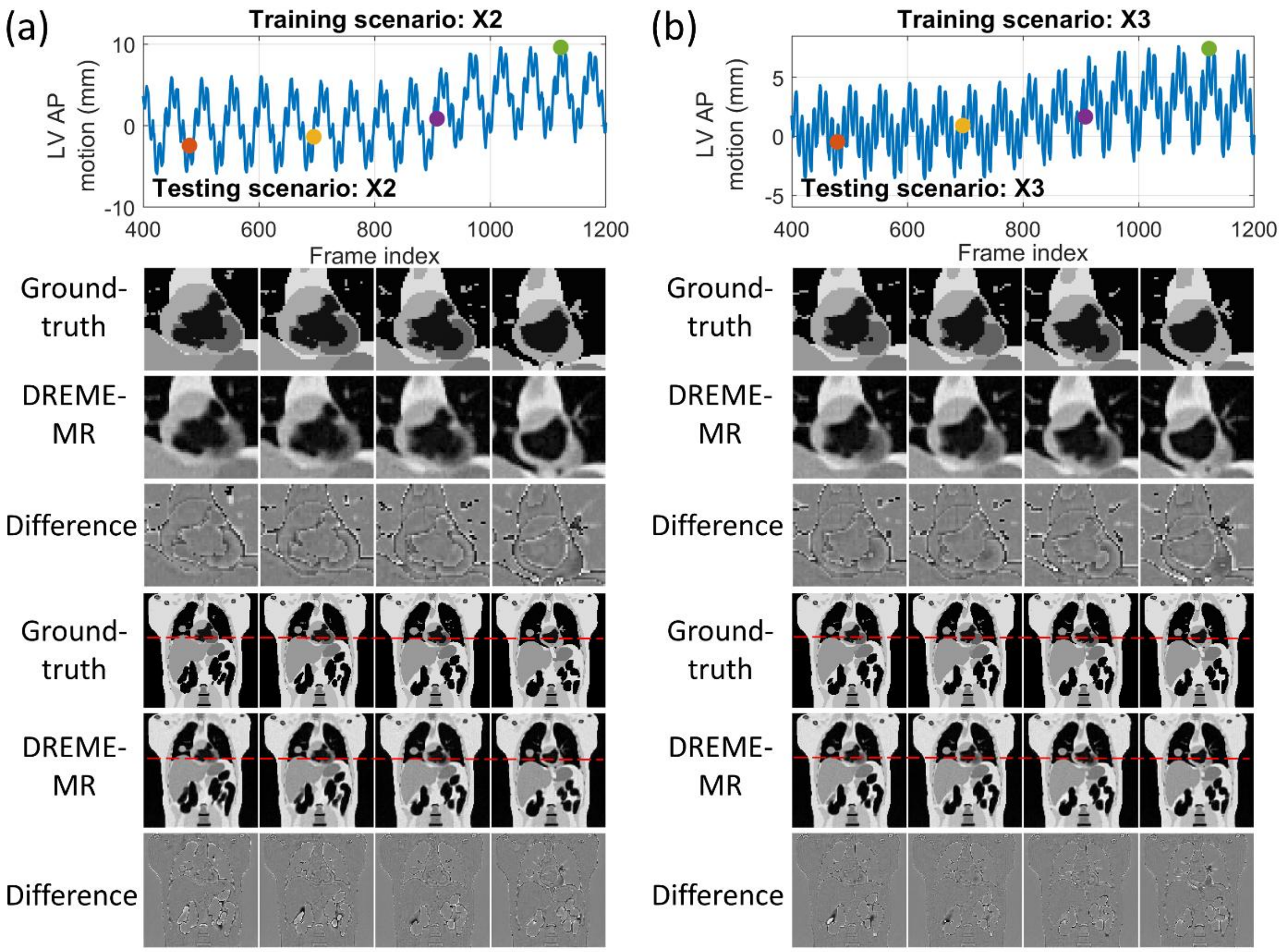
(a)
Training scenario: X2
Testing scenario: X2
LV AP motion (mm)
Frame index
(b)
Training scenario: X3
Testing scenario: X3
LV AP motion (mm)
Frame index
Ground-truth
DREME-MR
Difference
Ground-truth
DREME-MR
Difference

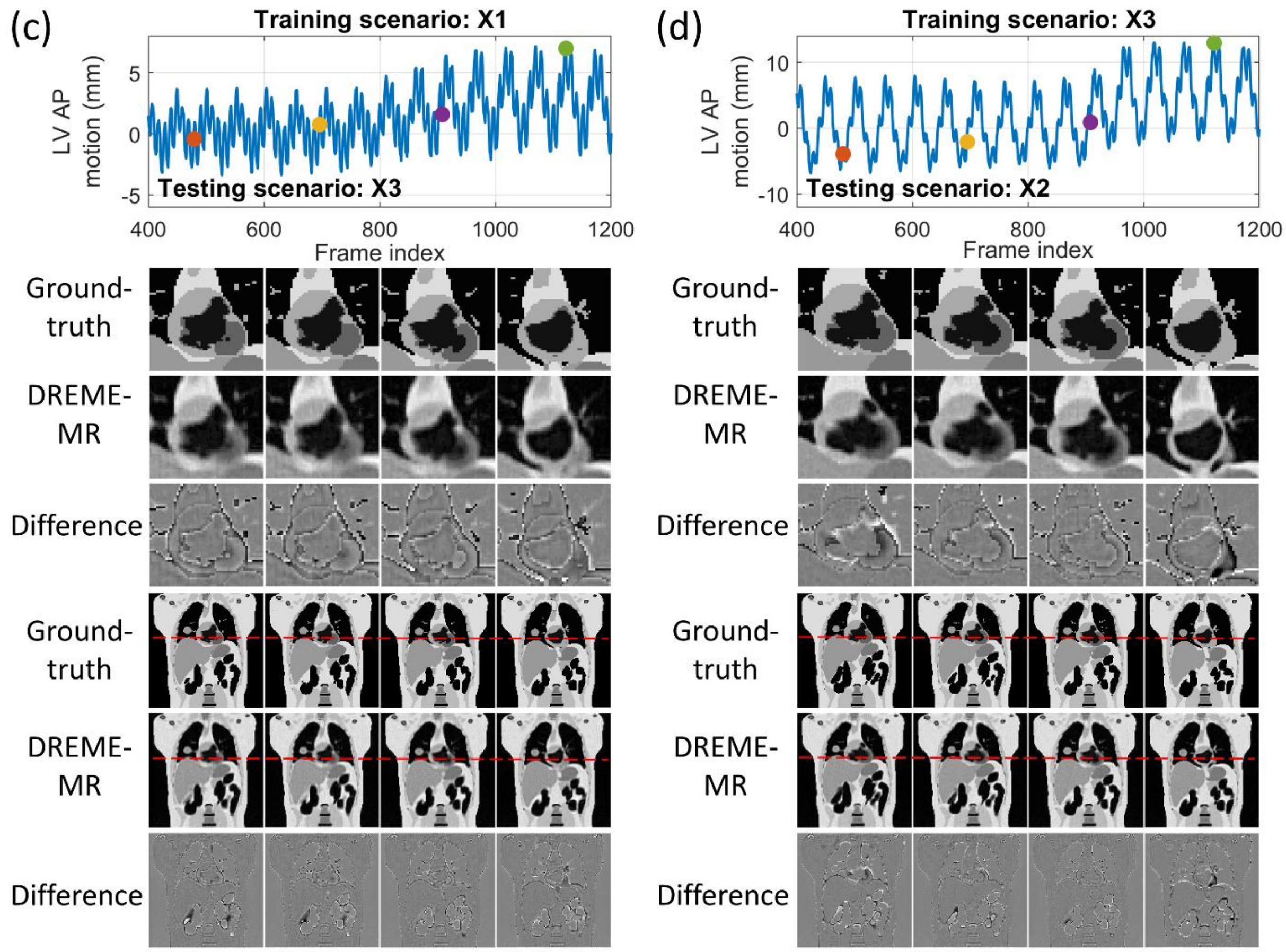

**Figure 4.** Representative examples of (a-b) dynamic reconstruction and (c-d) real-time imaging of the DREME-MR models in the XCAT study. The training and testing scenarios are labeled in each sub-figure. For each sub-figure, the first row shows the estimated LV motion curves along the AP direction, with the dots indicating the selected time points for plotting. The following rows compare the estimated MR images with the 'ground-truth' images at the four time points. Rows 2-4 are magnified views of the heart, corresponding to the full-volume images in rows 5-7. The window widths of the difference images are half of those of the MR images.

## 4.2 The human subject study

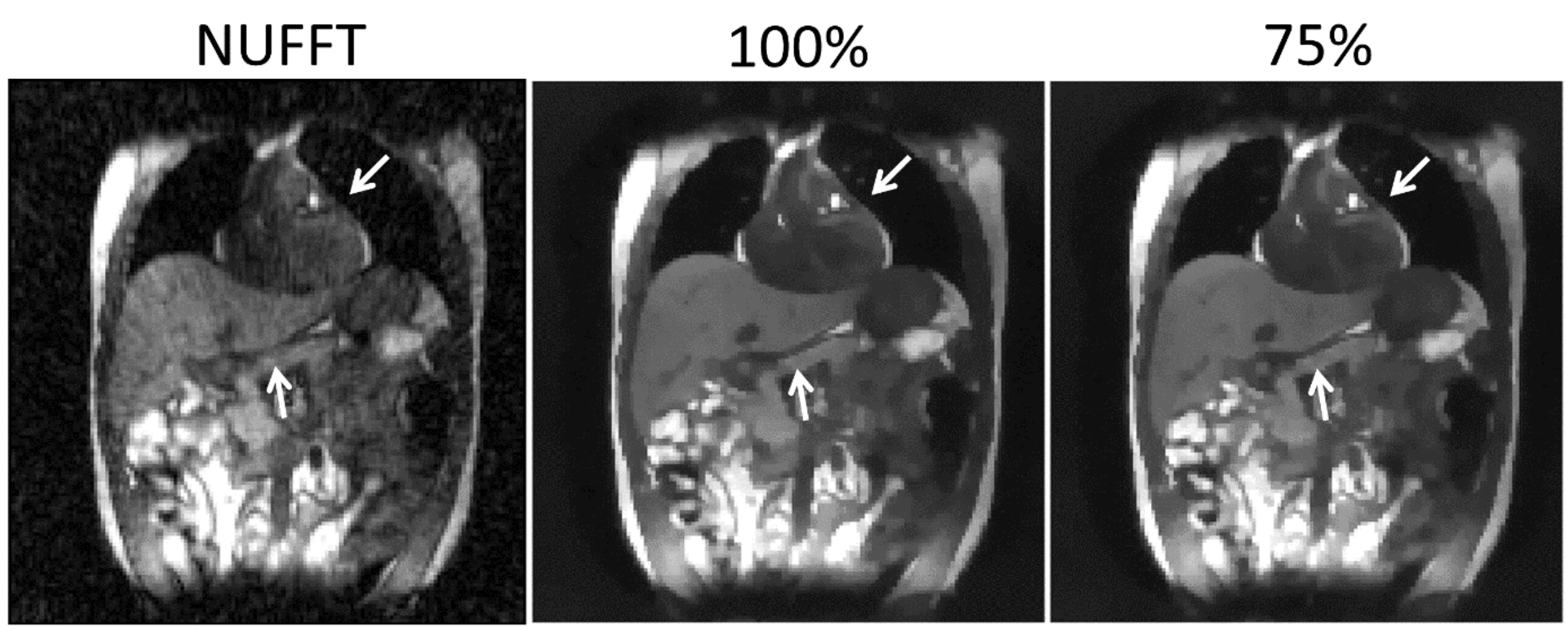

**Figure 5.** Comparison of reconstructed reference anatomy in the human subject study. The first panel shows the NUFFT-based reconstruction, using coil-compressed k-space data. The second and third panels compare DREME-MR-reconstructed reference anatomy using 100% and 75% of k-space data, respectively. Arrows highlight the areas with sharper anatomy.

Figure 5 compares MR images reconstructed using NUFFT and DREME-MR in the human subject study. Two DREME-MR models were trained, using 100% and 75% of k-space data (Sec. 3.5.2), as indicated by the figure titles. DREME-MR removed image noise and artifacts observed in the NUFFT reconstruction, and showed better-defined anatomical structures as highlighted by arrows. The match between the reference anatomy of the 100% and 75% models demonstrated that DREME-MR can successfully reconstruct a dynamic MRI set under a 220-s scan time (for the 75% model). Figure 6 compares the filtered liver and LV center-of-mass trajectories with motion surrogate signals extracted from the k-space data. The first and second halves of each panel correspond to the tasks of dynamic reconstruction and real-time imaging, respectively. Table 6 summarizes the Pearson correlation coefficients between the DREME-MR-resolved trajectories and the surrogate signals for the liver and the LV. Overall, high correlation coefficients (> 0.95) are observed for both tasks for liver motion tracking. In comparison, for the LV motion tracking, the correlation coefficient decreased to 0.63 and 0.34 in the left-right (LR) and AP directions, matching the observations from the XCAT study. Table 7 shows the DVF quality evaluation. Similar to the XCAT study, small SD*log*(J) values and negligible $J<0$ indicate smooth and physiologically plausible motion fields. Figure 7 presents the DREME-MR-resolved MRIs of the human subject, with both MRI volumes from the dynamic reconstruction (learning task 1) and from real-time tracking (learning task 2) shown.

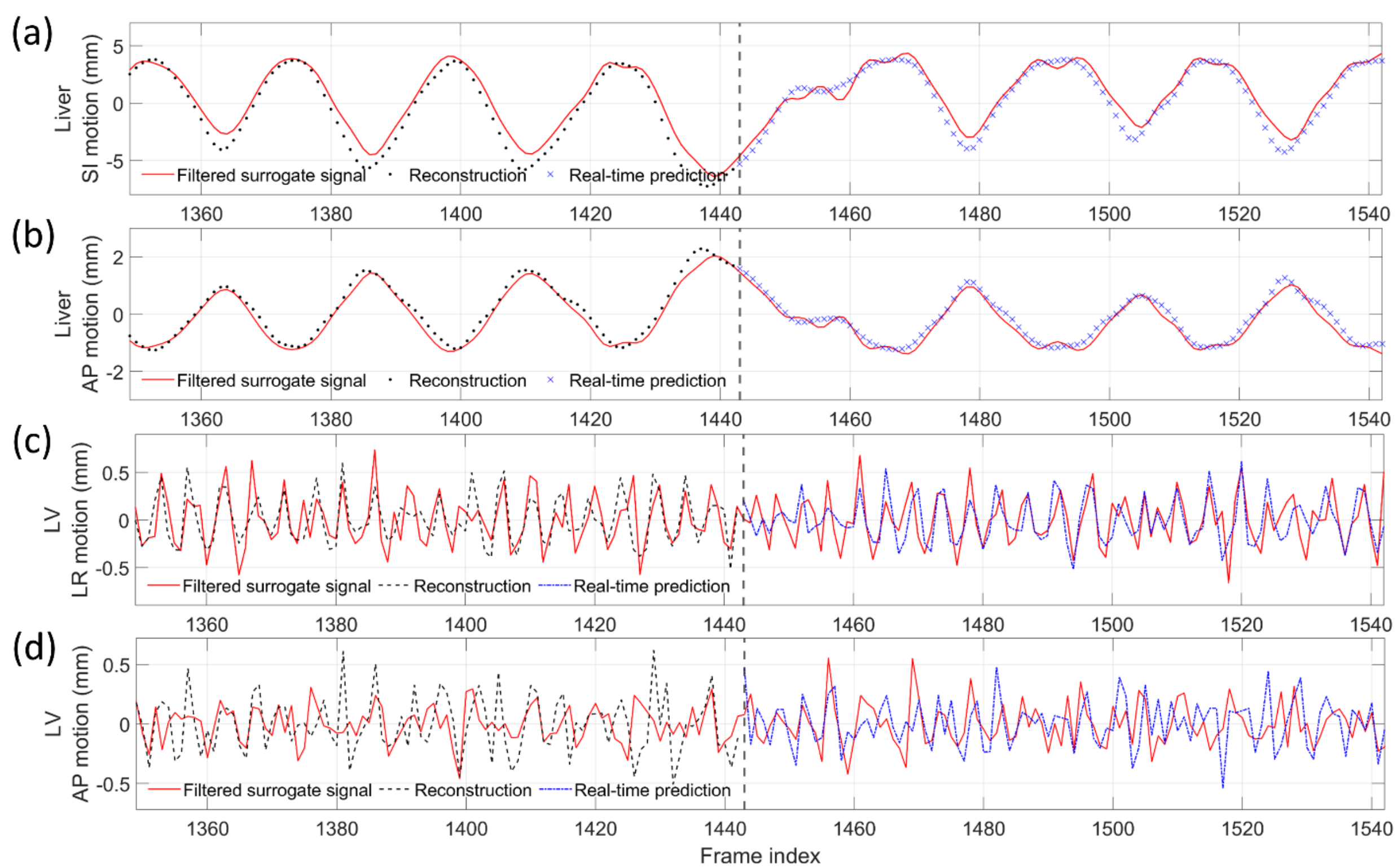

**Figure 6.** Comparison of DREME-MR-estimated liver and LV motion curves with the motion surrogate signals of the human subject study. (a-b) show the liver center-of-mass trajectories in the SI and AP directions. (c-d) show the LV center-of-mass trajectories in the LR and AP directions. The surrogate signals are extracted from the zero-frequency components of the k-space data. The high- and low-frequency components are filtered out from the curves to emphasize respiratory and cardiac motions of the liver and LV, respectively. The vertical dashed lines at frame 1,443 show the separations between dynamic reconstruction and real-time imaging tasks. Due to space constraints, only results corresponding to frame indices 1350 to 1542 are displayed.

**Table 6.** Pearson correlation coefficients of the liver and the LV, between surrogate signals curves and the DREME-MR-resolved motion trajectories.

| Task | Pearson correlation coefficient | | | |
|---|---|---|---|---|
| | Liver SI | Liver AP | LV LR | LV AP |
| Dynamic reconstruction | 0.997 | 0.950 | 0.634 | 0.341 |
| Real-time imaging | 0.986 | 0.987 | 0.695 | 0.358 |

**Table 7**. Deformation vector field quality evaluation for the healthy subject study. The standard deviation of the logarithm of the Jacobin determinant (SD*log*(J)) and the percentage of the voxels with negative Jacobian determinants (J<0) are evaluated to assess the smoothness and physiological reality of the motion fields. The results were reported as mean±SD.

| Task | SD*log*(J) | J<0 (%) |
|---|---|---|
| Dynamic reconstruction | 0.024±0.006 | 0.000±0.000 |
| Real-time imaging | 0.026±0.006 | 0.000±0.000 |

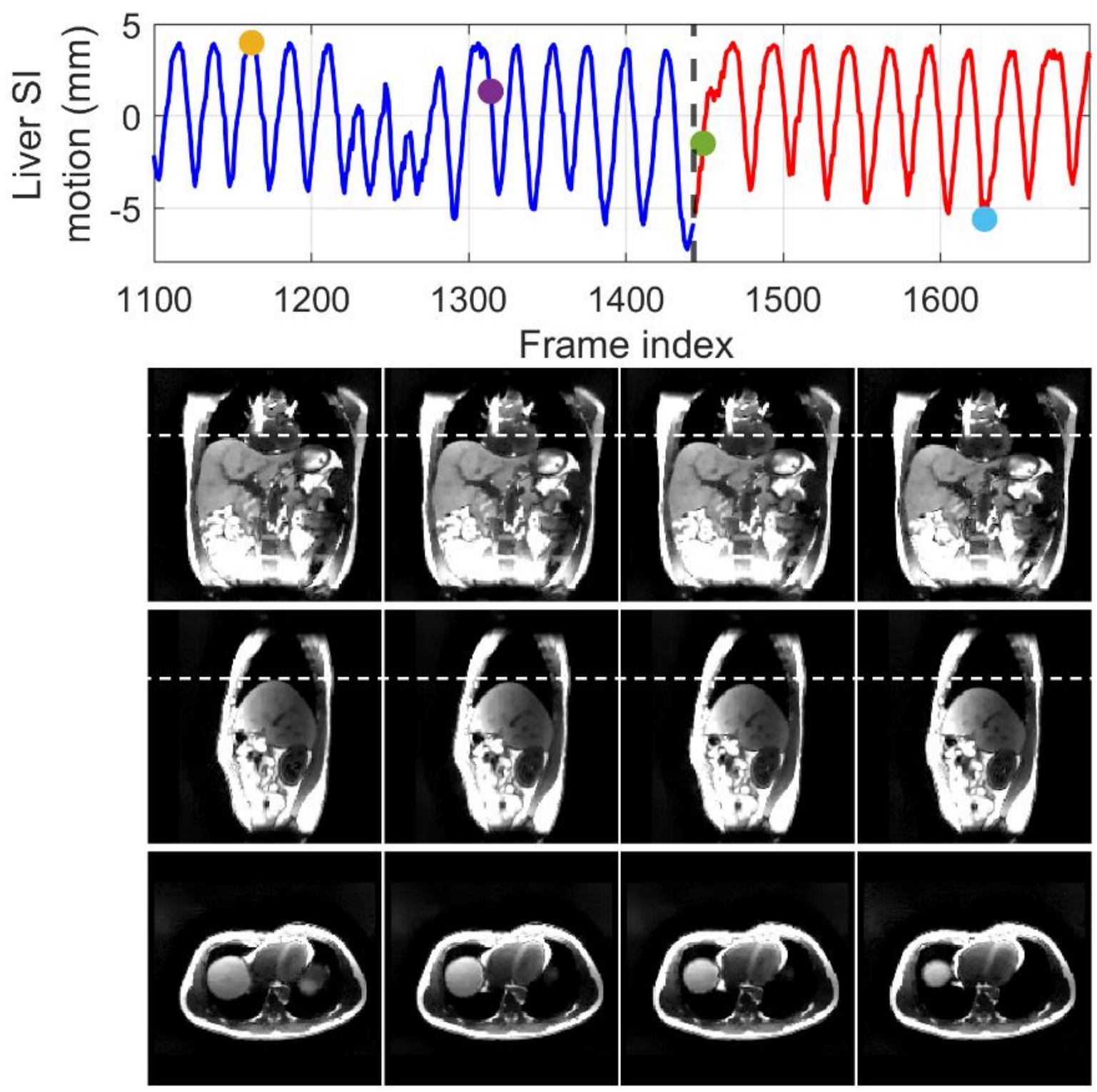

**Figure 7.** Dynamic MR images of the human subject study. The first row shows the estimated liver center-of-mass trajectory along the SI direction, with dots indicating the time points selected for plotting. The vertical dashed line at frame 1,443 shows the separation between the dynamic reconstruction and real-time imaging tasks. The subsequent rows display the selected MR images in the coronal, sagittal, and axial views. The first two columns correspond to the dynamic reconstruction task, while the last two columns correspond to the real-time imaging task.

## 5. Discussion

In this work, we proposed a dual-task learning framework, DREME-MR, for dynamic MRI reconstruction and real-time motion estimation. Clinically, dynamic MRI offers rich anatomical information for motion characterization, enabling personalized motion management strategy development and optimization, while real-time imaging and motion tracking enable real-time treatment adaptation and dose verification. DREME-MR adopts a 'one-shot' learning strategy, without requiring an external dataset for pre-training. The dual-task learning design offers several conceptual and practical advantages: (1) It eliminates potential biases introduced from patient-specific prior knowledge to train a model for real-time imaging. (2) It directly integrates the most up-to-date anatomical and motion information learned from dynamic MRI reconstruction into the real-time imaging process, thus minimizing the uncertainty in patient anatomy and motion tracking. (3) It unifies two interrelated tasks into a single framework, thereby streamlining clinical workflow. With its high spatiotemporal resolution (3 mm and 100-150 ms), DREME-MR has the potential to resolve both respiratory and cardiac motion.

DREME-MR was validated using an XCAT simulation study and further tested on a healthy human dataset. The XCAT results demonstrated the effectiveness of the 'one-shot' learning strategy, capturing various irregular respiratory motion patterns in dynamic and real-time MRI reconstruction. Cross-testing results showed that DREME-MR can generalize to unseen real-time motion scenarios, as the MLP-based motion encoder successfully extrapolated to unseen motion patterns. In the human subject study, DREME-MR achieved high Pearson correlation coefficients with extracted motion surrogate signals for respiratory motion. However, the correlation coefficients decreased for LV motion tracking. The total latency for real-time target localization was < 165 ms (= 100-150-ms data acquisition + 15-ms inference time). These preliminary results show a promising framework for real-time MR-guided adaptive radiotherapy.

Comparing the tracked respiratory motion and cardiac motion (Figures 3 and 6), it can be observed that the respiratory motion is dominant and much more significant than the cardiac motion. It echoes the previous reports that the average ratio between respiratory and cardiac excursions is approximately 11:1 (Petzl et al. 2024). Especially for the SI direction, the cardiac motion component is limited in the overall LV motion curve (Figure 3). For the AP direction, where the respiratory motion is less dominant, the high-frequency cardiac motion can be better visualized. For the patient study, due to the lack of 'ground truth', we used motion surrogates directly extracted from the k-space center as a reference for comparison with DREME-MR-resolved motion trajectories (Figure 6 and Table 6). The results of the human subject study demonstrate that DREME-MR can estimate respiratory motion in a real clinical environment, with high Pearson correlation coefficients for the liver in both the SI and AP directions (0.950-0.997) (Table 6). In comparison, the correlation coefficients for the LV, with a focus on the cardiac motion, dropped to 0.634-0.695 in the LR direction and 0.341-0.358 in the AP direction. Besides the difficulty in resolving cardiac motion from the dominant respiratory motion, a potential cause of the lower correlations is that the MR pulse sequence for the human scan was not optimized for cardiac imaging. As shown in Figures 5 and 7, the MR images show strong intensity from body muscle and fat, while the heart exhibited relatively low contrast. As our motion model is driven by motion-induced image contrast variation, this imbalance in intensity distribution results in a bias favoring respiratory motion. A potential future direction is to optimize the pulse sequence to enhance heart image contrast using our in-house scanners. For example, the 3D fast interrupted steady-state sequence offers high blood-muscle contrast-to-noise ratio images while still maintaining sufficiently short TRs for real-time imaging. This sequence was previously applied in free-running whole-heart MRI on both 1.5T and 3T scanners (Bastiaansen et al. 2020). In addition, the surrogate signals extracted for cardiac motion may also be more error-prone and less reliable than those for respiratory motion, as the cardiac motion is more localized, has a smaller magnitude, and is more susceptible to high-frequency noise, making it more difficult to extract reliably from the k-space center. In particular, the correlation coefficient for LV tracking increases for both the LR and AP directions if the DREME-MR model trained on the 75% k-space data was compared against the one trained on 100% of the data when evaluating the real-time prediction accuracy on the last 25% of data. For example, the value increased from 0.695 (vs. cardiac motion surrogate) to 0.849 (vs. 100% training data) in the LR direction. This increase indicates the potential uncertainty of the surrogate signals and self-consistency of DREME-MR in resolving cardiac motion. Alternative verification strategies include using the tagged MRI technique (Dornier et al. 2004) as an independent image-based verification or using electrocardiographs concurrently acquired during MRI scans as the cardiac surrogate signals (Thompson and McVeigh 2006), which provide more accurate and reliable cardiac motion signals.

Study results indicate that all DREME-MR variants exhibited comparable image quality (Table 2). However, the motion tracking results (Table 3) show more differences. While all variants had comparable tracking accuracy in dynamic reconstruction, DREME-MR outperformed the other variants in LV localization accuracy in real-time imaging. This indicates that the order of our sequential registration can better decouple and describe cardiorespiratory motion. Compared with MR-MOTUS (Huttinga et al. 2022), DREME-MR simultaneously optimizes the image and motion model to enhance the coherence and consistency while minimizing imaging artifacts. DREME-MR also eliminates the need for surrogate signals and corresponding motion sorting/binning. Using only the k-space data acquired from each pre-treatment MR scan, DREME-MR learns the patient anatomy and builds a motion model in a purely data-driven manner through a ‘one-shot’ learning strategy, without relying on an external large dataset for model pre-training and thus not susceptible to generalizability issues of conventional DL models. Since DREME-MR reconstructs the latest 3D anatomy and motion model based on the pre-treatment MR scan, which is immediately acquired prior to each radiation treatment delivery, it eliminates the uncertainties from day-to-day motion variations and anatomy changes and effectively avoids the biases from patient-specific prior knowledge encountered in registration-based deep learning methods (Terpstra et al. 2021, Nie and Li 2022, Shao et al. 2022). Then, based on the learned anatomy and motion model, DREME-MR can quickly and continuously infer real-time volumetric motion and MRIs from limited k-space signals to guide radiotherapy treatments, using the motion encoder optimized from the second learning objective. Compared with MRSIGMA (Wu et al. 2023), DREME-MR does not require pre-computing a motion dictionary derived from motion-sorted 4D-MRI. Instead, the MLP-based motion encoder directly learns the correlation between MR signals and dynamic motion states. Therefore, DREME-MR can adapt to a broader range of motion patterns, making it more robust to irregular motion.

In the XCAT study, motion tracking accuracy in real-time imaging was slightly lower than in dynamic reconstruction (Table 3), which is expected due to the unseen motion variations in the real-time imaging scenarios. We found that introducing a second coordinate system and applying frequency-domain regularization effectively decoupled respiratory and cardiac motion in the dynamic reconstruction task. In comparison, when DREME-MR was cross-tested on other motion scenarios in real-time imaging, Fourier analysis revealed a greater presence of respiratory frequency components in the cardiac MBC scores $\mathbf{w}_{\mathrm{i}}^{\mathrm{c}}(\mathrm{t})$, likely contributing to errors in the estimated cardiac motion. A potential solution is deformable augmentation, which may help disentangle cardiac and respiratory signals in $\boldsymbol{s}(\boldsymbol{k}, t)$ for unseen motion scenarios. The deformable augmentation resamples the learned MBC scores $\boldsymbol{w}_i(t)$ during training to synthesize anatomies with augmented respiratory and cardiac motion, enabling the motion encoder to generalize better and reduce overfitting. We previously implemented this strategy in our DREME framework for X-ray imaging and found it was effective (Shao et al. 2025). However, when applied to this study, deformable augmentation did not improve results in the human subject study. We found a potential cause is that inaccuracies in the coil sensitivity map may lead to inconsistencies in k-space signal synthesis, limiting its effectiveness. As a result, we did not include this strategy in the current work. To address this issue, we are curating an in-house dataset to further investigate the underlying causes and develop improved strategies.

Another limitation of DREME-MR is its training time. Currently, model training takes approximately 200 minutes on an NVIDIA Tesla V100 GPU. A potential approach to accelerate dual-task learning is to use a more efficient anatomical representation. Recently, 3D Gaussian representation (Fei et al. 2024) has been applied in medical image reconstruction, primarily for X-ray-based CT/CBCT reconstruction. In this approach, anatomy is represented as a collection of Gaussian distributions whose attributes, such as position, orientation, and size, are learnable parameters. Compared to voxel-based representations, 3D Gaussian representation has been shown to provide a sparse and efficient volumetric representation of the human anatomy. It is expected that adopting a Gaussian representation could significantly reduce computation time, though this remains to be further investigated. In addition, the reference volume and motion model of DREME-MR can potentially be pre-conditioned or meta-learned with patient-specific priors or population-based data, and then fine-tuned subsequently using patient-specific acquisitions for further reconstruction acceleration.

## 6. Conclusion

In this study, we proposed DREME-MR, a dual-task learning framework for dynamic MRI reconstruction and real-time motion estimation. DREME-MR achieved overall accurate respiratory and cardiac motion tracking in the XCAT simulation study, although cardiac motion tracking was found to be more challenging than respiratory motion tracking, especially for the respiration-dominant superior-inferior direction. For the human subject study, DREME-MR demonstrated high liver motion correlations with surrogate signals but moderate LV motion correlations, likely due to additional challenges from the suboptimal pulse sequence for cardiac imaging and a lack of reliable cardiac motion surrogates. DREME-MR represents a promising step toward real-time MRI-based motion tracking for MRI-guided radiotherapy.

## Acknowledgements

The study was supported by funding from the National Institutes of Health (R01 CA240808, R01 CA258987, R01 CA280135, R01 EB034691), and from Varian Medical Systems. We would like to thank Dr. Paul Segars at Duke University for providing the XCAT phantom for our study.

## Ethical statement

The healthy human subject dataset is publicly available and fully anonymized. No ethical approval was required. This is a retrospective analysis study and not a clinical trial. No clinical trial ID number is available.